\documentclass[fleqn]{aa}    
\usepackage{graphicx}
\usepackage{graphics}
\usepackage{txfonts}
\usepackage{supertabular}
\begin{document}
   \title{ Planetary nebulae in the Magellanic Clouds~:
   \thanks{Based on observations made at the European Southern Observatory, La~Silla, Chile}
\fnmsep
   \thanks{Tables 3,4,5,6,A1 and A2 are only available in electronic form
at the CDS via anonymous ftp to cdsarc.u-strasbg.fr (130.79.128.5)
or via http://cdsweb.u-strasbg.fr/cgi-bin/qcat?J/A+A/}
            }
   \subtitle{ II) Abundances and element production}   
   \author{P. Leisy \inst{1, 2}
   \and    M. Dennefeld \inst{3}
          }
\offprints{P. Leisy \email{pleisy@ing.iac.es}}
 \institute{Isaac Newton Group of telescopes ING/IAC, La Palma, SPAIN.
 \and  Part of this work was done while at ESO La Silla, CHILE and in Pr\'erebois, Rombach-le-Franc, FRANCE
 \and Institut d'Astrophysique de Paris, CNRS, and Univ. P. et M. Curie, 98bis B$^d$ Arago, F-75014 Paris, FRANCE.
        }
   \date{Received 1 Feb 2005 / Accepted 1 Jun 2006} 
   \abstract
{We present the second part of an optical spectroscopic study of
planetary nebulae in the \object{LMC} and \object{SMC}.
The first paper, Leisy~\&~Dennefeld~(1996), discussed
the CNO cycle for those objects where C abundances were available.
}
{In this paper we concentrate more on other elemental abundances 
(such as O, Ne, S, Ar)
and their implications for the evolution of the progenitor stars. 
} 
{ We use a much larger sample of 183 objects, of which 65 are 
our own observations,  where the abundances have been 
re-derived in a homogeneous way.
For 156 of them, the quality of data is considered to be satisfactory 
for further analysis.
}
{We confirm the difficulty of separating Type$\,$I and 
non-type-I objects in the classical He-N/O diagram, as found in Paper I, 
a problem reinforced by the variety of initial 
compositions for the progenitor stars.
We observed oxygen  variations, either depletion via the ON cycle in the 
more massive progenitor stars, or oxygen production  in other objects. 
Neon production also appears to be present. 
These enrichments are best explained by   
 fresh  material from the core or from burning shells, 
brought to the surface by the $3^{rd}$ dredge-up, as reproduced in recent 
models (some including overshooting). 
All the effects appear stronger in the \object{SMC}, suggesting a higher 
efficiency in a low metallicity environment. 
}
{Neither oxygen nor neon can therefore be used to derive the initial composition
of the progenitor star: 
other elements not affected by processing such as sulfur, argon or, if observed,
chlorine, have to be preferred for this purpose. 
Some objects with very low initial abundances are detected, but on average, 
the spatial distribution of PNe abundances is consistent with the history of 
star formation (SF) as derived from field stars in both Clouds.
}
 \maketitle

\section{Introduction}
Planetary nebulae (PNe) are prime targets for
studing the chemical evolution of nearby galaxies. This is particularly 
true for the two Magellanic Clouds, which have a different composition  
from our own Galaxy, thus allowing the study of metallicity effects.
Their distances are known with reasonable accuracy, and thus 
the physical properties of the PNe can be determined easily; and they are 
close enough that a large number of PNe have already been identified,
therefore giving statistical significance to the derived properties.

In our first paper (Leisy~\&~Dennefeld~(\cite{leis2}); hereafter Paper~I) we 
reported  our observations of PNe in the Magellanic Clouds, and discussed 
those objects (16 in \object{LMC} and 15 in \object{SMC}) for which carbon
abundances were available.
The present paper discusses the entire sample for which optical observations 
are available: 65 objects from our own observations, to which we add  
118 other PNe already published in the literature. All the data were 
treated for the first time in a homogeneous way, 
allowing a global analysis of elemental abundances  in the Magellanic Clouds. 

\noindent
 At the end of their evolution, intermediate-mass stars pass
through an  asymptotic giant branch (AGB) phase and experience helium
shell flashes (Thermal-Pulse AGB).
During these  phases, they  lose most of their envelopes.
These envelopes contain both original, un-processed  material 
and  material synthesized
in the deep layers of the star, and then  dredged-up during the Red Giant
and AGB phases.
Therefore substantial mass is ejected, which enriches  
the interstellar medium (ISM) with processed elements such as He, C, N 
and s-process elements (see for
example Renzini~\&~Voli~\cite{ren1}, Iben~\&~Renzini~(\cite{iben3}),
Dopita~\&~Meatheringham~(\cite{dop2}), Vassiliadis \&~Wood~\cite{vasi2}). But
 the ISM is also enriched in other, un-processed elements that were 
 incorporated from the ISM at the time  
 the progenitor star was formed, such as O, S, Ar, or Fe, which are
elements believed to remain un-processed during the evolution of the PN 
progenitor star. 


 For those elements synthesized during the lifetime of their  
progenitor stars, the study of PNe 
allows one to determine how the production of chemical elements
depends on initial mass or metallicity and how the products are
{\it dredged-up} during the late stages of stellar evolution.
Although Kaler,~Iben~\&~Becker~\cite{kal1}, Becker \&~Iben~(1979, 1980),
Renzini~\&~Voli~\cite{ren1}, or Sweigart,~Greggio,~Renzini~(1989, 1990)
have demonstrated reasonable agreement
between observations and the dredge-up theory for giant stars, 
quantitative discrepancies still exist.
We showed in Paper~I that 
initial metallicity  plays a crucial role in the efficiency of the 
various reactions and that additional processes, such as Hot Bottom Burning, 
predicted by theory (Gronewegen and de Jong,~\cite{groe}) were effectively observed.
Recent  evolution models of intermediate mass stars incorporating this are now 
available for a direct comparison with observations 
(Marigo et al.~(\cite{mari1}), Marigo~(\cite{mari2}) or Marigo et al.~\cite{mari3}).
 
While PNe  play an important role in the chemical evolution of
their host galaxies by furnishing helium, carbon, and nitrogen in
great quantities, they can also be used to trace the evolution of those 
elements not affected by processing in their progenitor stars. 
With masses in
the range $1~M_{\sun}~\leq~M_{i}~\leq~8~M_{\sun}$, evolution can be traced from PNe
over the past 10 billion years, provided the mass of the central star and 
its initial metallicity are known. The latter is usually derived from 
oxygen, but in view of the metallicity effects shown in Paper~I and the 
involvement of O in the CNO cycle (as discussed later in this paper), 
other heavy elements, such as neon, sulfur and argon will also have to be used. 
Present day abundances in the ISM,
used for comparison, are determined from H\,{\sc ii}~regions. \\
A large sample of objects, with various metallicities and
an homogeneous data reduction and determination of abundances
such as the one  provided here, 
is therefore crucial to understand the role of the various
parameters involved. It will also allow a global analysis of the 
chemical evolution of the Magellanic Clouds
to be compared with the results obtained  from 
H\,{\sc ii}~regions (e.g. by Dufour~\cite{duf1},
20 objects) and with those obtained from direct studies of a few 
stellar spectra (Spite et al.~\cite{spit1}; Hill et al.~\cite{hill1}).
Such a program is a long-term observational enterprise, 
and first results have already been
presented in various conferences (e.g. in Leisy \& Dennefeld~\cite{leis3}).

A short description of the observations and data reduction is given again 
in Sect.~2, as a reminder from Paper~I.
The physical parameters for the objects, the abundances
determinations, and the corresponding uncertainties are given in Sect.~3.
Readers not interested in the details of the derivation of the PNe  
parameters  can go directly to Sect.~4, where the abundance patterns 
are discussed element by element. 
Implications for the chemical evolution of the Magellanic Clouds are 
presented in Sect.~5.
\section{Observations and data reduction}
The objects observed and named according to the SIMBAD nomenclature 
were mainly selected from the list of 
Sanduleak et al.~\cite{san1} (SMP LMC or SMP SMC), 
with a few additions from Henize~(\cite{heni}: LHA-120 Nxx for LMC, 
LHA-115 Nxx for SMC), 
Morgan \&~Good~(\cite{mor2}: MGPN LMC N=1,86),
Morgan~(\cite{mor3}: $[M94b]$ N=1,54),
Morgan \&~Good~(\cite{mor1}: MGPN SMC N=1,13) and
Morgan~(\cite{mor4}: $[M95]$ N=1,9).

Our spectra were taken during different observing runs from 1984 to 1999 with
various telescopes and instruments in ESO La Silla~: the
1.52m, 2.2m, 3.60m and NTT telescopes, and the 
B\&C, EFOSC, EMMI spectrographs. 
The spectral range was always split into two regions, blue and red,
with an overlap including at least the $H_{\beta}-[OIII]$ lines.
This choice was made to ensure an adequate spectral resolution
(about 3-4 \AA/pixel{\bf )} and yet provide adequate spectral coverage.
All spectra were reduced using  MIDAS and its LONG package, following 
a standard procedure for 2D spectra, with 
 optimized spectrum extraction and line fitting to 
measure the line intensities.
 Details of the observing and data reduction procedures can be found
in Paper~I (Sect.~2 and 3).

To our own observations we added data for other objects taken from the 
literature, namely Monk et al.~\cite{monk} (hereafter MBC),
Meatheringham and Dopita~(\cite{mea1},~\cite{mea2}, hereafter MD), and 
Vassiliadis et al.~(\cite{vasi1}).
We used corrected line fluxes from those lists (taking fluxes  corrected
for extinction as determined in the original paper) and then determined
new abundances with our code, to obtain a homogeneous set of determinations.
\label{sampleREMOVE}
The total sample is thus composed of 65 objects observed by us, to which we 
add 118 other objects (not in common)  with published line intensities.

\begin{figure}[htbp]
   \includegraphics[angle=-90,width=9.5cm]{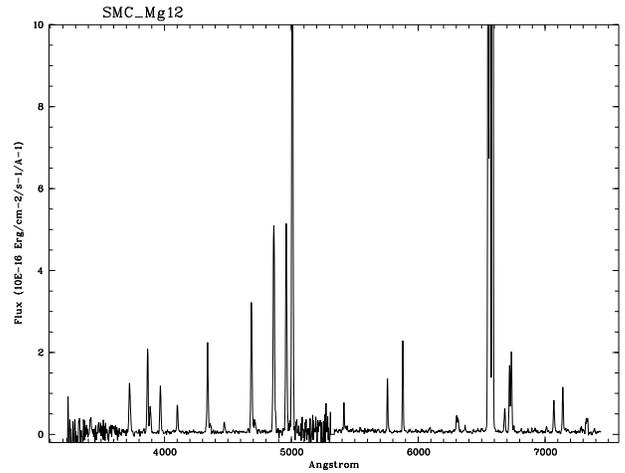}
   \caption{An example of a Type\,I PN in the \object{SMC} (MGPN SMC 12) 
with strong [NII] lines.}
   \label{MGPN-SMC-12}
\end{figure}

\begin{figure}[htbp]
   \includegraphics[angle=-90,width=9.5cm]{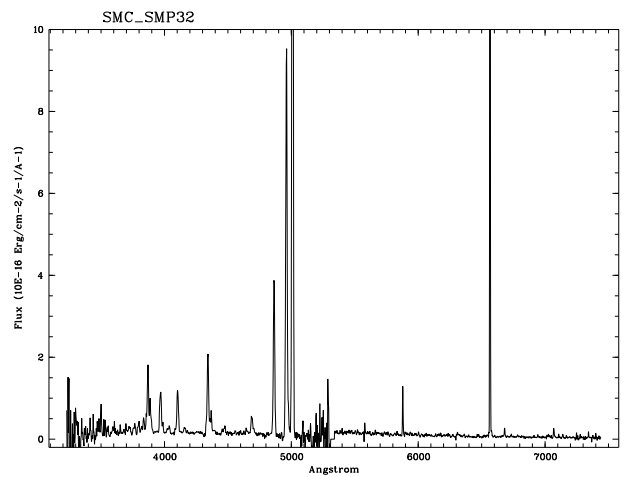}
   \caption{A standard faint PN in the \object{SMC} (SMP SMC 32). 
Note the absence of [NII] lines. }
   \label{SMP-SMC-32}
\end{figure}

\begin{figure}[htbp]
   \includegraphics[angle=-90,width=9.5cm]{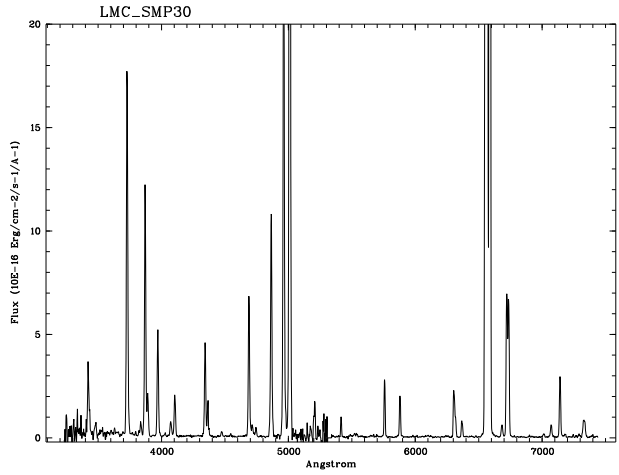}
   \caption{A bright Type\,I PN in the \object{LMC} (SMP LMC 30) with
many emission lines observed, including  [NeV] in the near UV.}
   \label{SMP-LMC-30}
\end{figure}

\begin{figure}[htbp]
   \includegraphics[angle=-90,width=9.5cm]{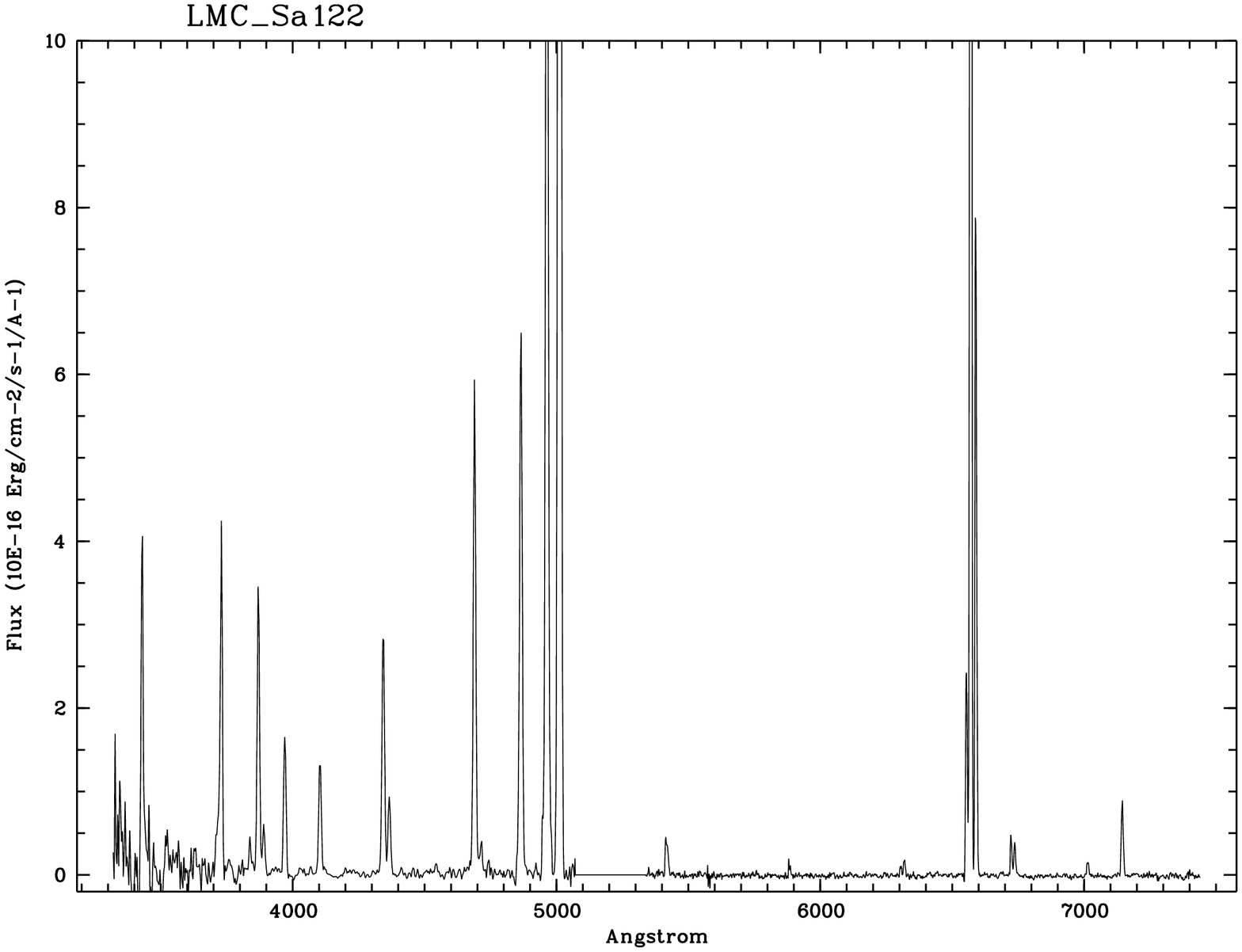}
   \caption{A Type\,i  PN in the \object{LMC} (SMP LMC 122)}
(strong [NII] but very weak He\,I lines )
   \label{SMP-LMC-122}
\end{figure}

\begin{figure}[htbp]
   \includegraphics[angle=-90,width=9.5cm]{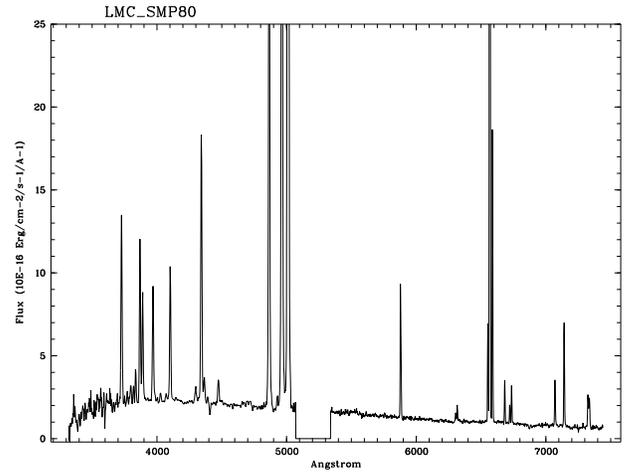}
   \caption{A standard PN in the \object{LMC}    (SMP LMC 80). 
 The gap between the blue and red spectra is produced by the dichroic
in the  EMMI spectrograph at  the NTT, used to record both 
wavelength ranges  simultaneously.}
   \label{SMP-LMC-80}
\end{figure}

To illustrate the high quality of our spectra, we show some examples
in  Fig.~\ref{MGPN-SMC-12},~\ref{SMP-SMC-32},~\ref{SMP-LMC-30},~\ref{SMP-LMC-122}, and ~\ref{SMP-LMC-80},  
of the PNe of various types in both Clouds. The blue and the red 
spectral ranges were observed separately, but merged in the 
illustrations. The S/N can be judged, 
for instance, from the strength of the [OIII]  $\lambda_{4363}$  line, 
a key line for temperature determinations.

As discussed in more detail in Sect.~4, a first analysis of the results 
showed some discrepant  points in the diagrams, some of which could clearly 
be ascribed to poor quality spectra for some objects from the literature. 
We thus removed from the initial sample of 183 objects 8 PNe in the 
\object{SMC} and 18 in the \object{LMC}, where  no [OIII] temperature 
determination was possible, 
and no lines fainter than 10\% of $H_{\beta}$  were detected. 
This leaves us with a sample of 157 objects for the full analysis. 


\subsection{Comparison of line intensities }
\begin{figure}[htbp]
   \includegraphics[angle=-90,width=9cm]{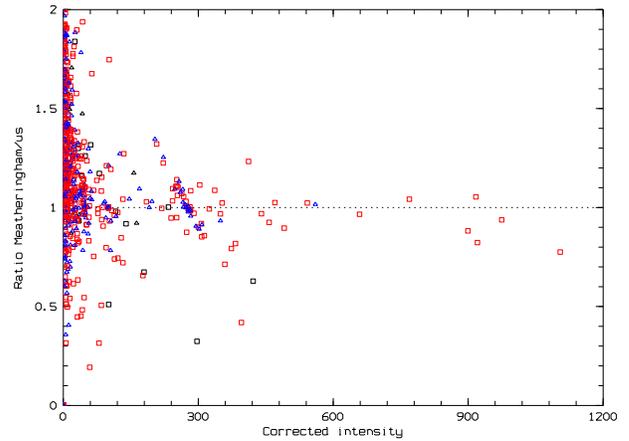}
   \caption{Ratio of line intensities in the two independent observations, 
for  the 21 PNe in common between us and  Meatheringham
and Dopita~\cite{mea1};~\cite{mea2}.}
   \label{lineCOMP1}
\end{figure}

We have 30 objects from our observed sample  in common with MD
and 28 with MBC, thus providing a possibility for consistency checks.
In Fig.~\ref{lineCOMP1} we present the comparison of the line intensities 
for the 19 \object{LMC} and 11 \object{SMC} objects in common 
with Meatheringham and Dopita~(\cite{mea1}) and Dopita~(\cite{mea2}).

\begin{figure}[htbp]
   \includegraphics[angle=-90,width=9cm]{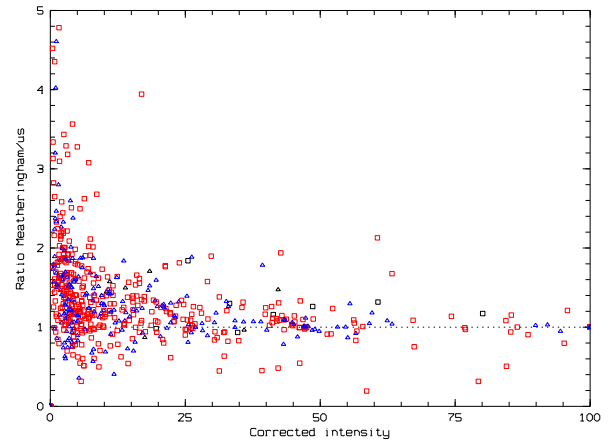}
   \caption{Zoom of the previous figure (Fig.~\ref{lineCOMP1}) 
at low intensities. }
   \label{lineCOMP2}
\end{figure}

At high intensities,  the agreement is generally good, with differences 
smaller than 20-30\%.
Most of the discrepancies come from the $[OII]_{3727}$\AA~ lines or from 
lines that are at the edge of the observed wavelength domain (as in Meatheringham and 
Dopita~\cite{mea1}, for instance), a difference that can most probably be 
ascribed to an imperfect extinction or to spectral response correction.
If we examine Fig.~\ref{lineCOMP2} in the low intensity line region, i.e. 
the intensities smaller than $H_{\beta}~\simeq~100$, we see larger discrepancies.
In particular MD always find higher intensities than we do for the faintest 
lines. This is a general trend that can probably be explained by a lower S/N
and poor determination of the continuum level. \\
We do not discuss the comparison in detail between our intensities and 
those of Monk et al.~\cite{monk} (28 objects in common)
or Vassiliadis et al.~\cite{vasi1} (11 objects in common), 
as   the quality of their data seems to be rather poor, as judged from 
the huge uncertainty/errors for those faint objects, and 
many of them have finally been rejected from the final analysis diagrams.
\section{Physical parameters}
\subsection{Extinction}
\label{sectionA}
The Whitford~(\cite{whit}) galactic extinction law (noted $f(\lambda)$) 
was used in the parameterized form of Miller and
Matthews~(\cite{milm}) for the optical domain. The uncertainties and differences 
between this galactic law and specific laws for the Clouds 
(e.g. Nandy et al.~\cite{Nand}) are such that this choice does not 
have a significant impact on the results in the optical range  
(contrary to the situation  in Paper~I, which included UV).
The Balmer theoretical ratios were then used with  ({\it
Eq.~\ref{fluxEQ}}) to determine the extinction coefficient 
$c(H_{\beta})$ in an iterative way, two iterations being necessary 
because densities
$n_{e^{-}}$ and temperatures $T_{e^{-}}$, which affect the expected ratios, 
were calculated at the same time:

\begin{equation}
\label{fluxEQ}
I_{cor}~(\lambda)/I_{cor}~(H_{\beta}) = 
I_{obs}~(\lambda)/I_{obs}~(H_{\beta}) 
10^{C(H_{\beta}) f(\lambda)}.
\end{equation}
Three ratios, $H_{\alpha}$/$H_{\beta}$, $H_{\gamma}$/$H_{\beta}$, and
$H_{\delta}$/$H_{\beta}$, were used to determine the extinction coefficient.
When the three ratios gave very discrepant values, generally because of 
a low S/N ratio in the fainter lines,
only the $H_{\alpha}$/$H_{\beta}$ ratio was used
(see Paper~I, Sect.~4.1 for more details).
In the very few cases where we obtained
a non-physical negative value for $c(H_{\beta})$,
the extinction coefficient was then set to zero.

The observed and corrected line
intensities, $I_{obs}~(\lambda)$ and $I_{cor}~(\lambda)$, relative to 
$H_{\beta}=100$, are given for all the objects in $Appendix~A$ .
The corresponding $H_{\beta}$ fluxes from the literature 
(Meatheringham et al.~\cite{meam}) are also given in 
$10^{-16}\;\mathrm{erg}\;\mathrm{cm}^{-2}\;\mathrm{s}^{-1}$ units, not
corrected for extinction.

\subsection{Temperature and density}
\label{secTempDens}
The plasma diagnostics and the ion abundances were derived by solving
the five-level statistical
equilibrium equations with a code kindly made available by G.~Stasinska
and modified by us to update some coefficients or add more lines.

If possible we always calculated temperatures and densities from our code.
The rules adopted for the final choice of temperatures are the following: 
\begin{itemize}
\item If no $T_{[NII]}$ determination is available, then we take 
$T_{[NII]}$ = $T_{[OIII]}$  if
$T_{[OIII]}$ is in the 8000\,- 30000\,K range; otherwise 12000\,K is adopted
  \item If  $T_{[OIII]}$ is not determined, but $T_{[NII]}$ is available;
so we adopt $T_{[OIII]}$ = $T_{[NII]}$, if
$T_{[NII]}$ is in the 8000\,-30000\,K range (otherwise, 12000\,K is adopted 
arbitrarily)
  \item If no temperature  determination is available at all, an arbitrary 
value of 12000\,K is adopted. 
\end{itemize}
\begin{figure}[htbp]
   \includegraphics[angle=-90,width=9cm]{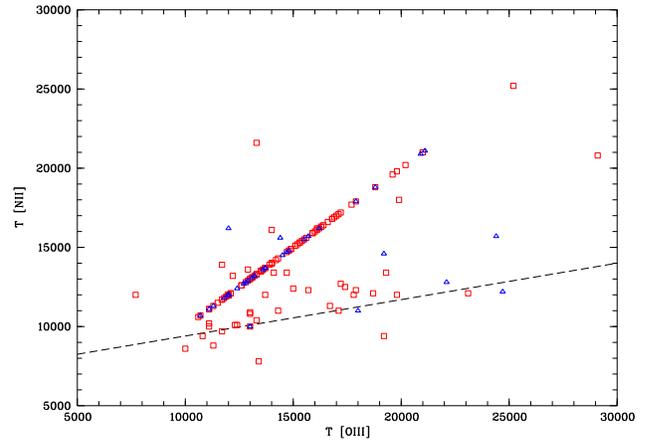}
   \caption{$T_{[NII]}$ versus the $T_{[OIII]}$ temperatures.  The relation
found by Kingsburgh \& Barlow~(\cite{king}) for Galactic objects is plotted as 
a dashed line. }
   \label{tempe}
\end{figure}

 We compare the two temperatures ( $T_{[OIII]}$ and $T_{[NII]}$) 
in Fig.~\ref{tempe}. 
The alignment of points along the slope 1 line is due to the adopted rule 
$T_{[NII]} = T_{[OIII]}$, as a first  approximation, when one of these 
temperatures could not be determined directly.  
Generally, however, both observations and models give $T_{[NII]}$ less
than $T_{[OIII]}$ (as seen in the same figure). 
A better approximation has been proposed for Galactic 
objects by Kingsburgh \& Barlow~(\cite{king}) but is not directly applicable 
here, because the  range of T 
and  $n_{e-}$ encountered is much larger, and 
 furthermore, such a relation is certainly very  metallicity dependent. 
The figure shows that it seems to be a good approximation for some  of the 
objects where  both temperatures are 
determined but that it generally underestimates  $T_{[NII]}$.

\noindent 
For electron density ($n_{e-}$) determinations, the rules adopted are 
the following.
We first derive  the densities from [SII] lines (if available, 
and if the density is less than 50000 $e^{-}.cm^{-3}$); otherwise,
the density is derived from the [ArIV] lines, when available. 
If neither of these determinations is possible, we adopt a 
density of 5000 $e^{-}.cm^{-3}$ (a value generally adopted in other published data).
More details of the treatment can again be found in Paper~I, Sect.~4.2.

\subsection{Chemical abundances-ionization correction factors}
Ionic abundances were calculated relative to hydrogen from the 
recombination lines of He or from the forbidden lines of heavier elements.  
Total elemental abundances were then derived through ionization
correction factors (ICF) to take the unobserved ionic stages into account.
The details of the procedure and the ICF used are described in Paper~I. 
The only change introduced in this paper is the use  of 
a more accurate ICF for argon (Eq.~\ref{EQicfAr1} and \ref{EQicfAr2}),
derived from models by Ratag \& Pottasch~(\cite{rata}) and based on 
 excitation class (EC)  and  observed ionic stages of the object.

When the [Ar\,III] and [Ar\,IV] lines are available, the ICF is~:
\begin{eqnarray*}
  ICF & = 1.39 & \qquad EC < 4.5 \nonumber \\
  ICF & = 1.43 & \qquad 4.5 < EC < 7.5 \nonumber \\
  ICF & = 1.75 & \qquad EC > 7.5
\end{eqnarray*}
\begin{equation}
\label{EQicfAr1}
\frac{N(Ar)}{N(H)} =  \frac{N(Ar^{2+})+N(Ar^{3+})}{N(H^{+})} * ICF(Ar)
\end{equation}
When only the [Ar\,III] line is observed, then~:
\begin{eqnarray*}
  ICF & = 1.73 & \qquad EC < 4.5 \nonumber \\
  ICF & = 2.04 & \qquad 4.5 < EC < 7.5 \nonumber \\
  ICF & = 3.85 & \qquad EC > 7.5
\end{eqnarray*}
\begin{equation}
\label{EQicfAr2}
\frac{N(Ar)}{N(H)} =  \frac{N(Ar^{2+})}{N(H^{+})} * ICF(Ar)
\end{equation}
The excitation classes are derived analytically with
the equations of Dopita~\&~Meatheringham~(\cite{dop1}).

\subsection{Uncertainties}
The final uncertainty on a given line depends primarily on its
strength and on the quality (S/N) of the spectrum.
The continuum noise level is a good indicator, but uncertainties in 
fitting  the lines are also taken into account.
The $H\beta$ flux uncertainties given in 
$Appendix~A$ 
are a good representation of the quality of the spectrum.

 For  lines with intensities down to about 1\% of  $H\beta$ (5\% for very 
faint objects), the uncertainty scales approximately with the inverse 
square root of their intensity ratio to $H\beta$.
 For very faint lines (fainter than the above limit), 
the S/N gets worse, and the uncertainty  scales approximately
with the inverse of their intensity ratio to $H\beta$.
As expected, for fairly bright objects, the majority of the spectra have 
very good S/N ratios, so they are among the best spectra available 
in the whole sample.
For the faintest objects, sometimes only a few bright lines are
well observed, and the quality of the results scales accordingly.
An estimate of the uncertainty in the extinction coefficient, c($H_{\beta}$), 
is also given, derived from the comparison of individual values obtained 
from each of the individual Balmer line ratios used in the determination. 
In addition to "absolute" uncertainties discussed above, comparing
abundances derived by various authors for objects in common provides another 
way to derive the final uncertainty of abundances.
  \begin{table*}[htb]
  \setlength{ \tabcolsep}{0.6mm}
  \caption{Abundance differences between us and Meatherigham et al.~(\cite{meam}).
The first line gives the  mean difference averaged over several objects 
(the difference can be higher for an individual one), the second line gives the 
number of objects used in the determination,  while the $\sigma$ is given 
in the third line.}
  \label{CompMeat}
  \begin{tabular}{|l|l|l|l|l|l|l|l||l|l|l|l|l|l|l|}
  \noalign{ \smallskip}
  \hline \noalign{ \smallskip}
   & \multicolumn{7}{|c|}{ \bf \object{LMC}} &  \multicolumn{7}{|c|}{ \bf \object{SMC}} \\
  \hline \noalign{ \smallskip}
  \hline \noalign{ \smallskip}
  \noalign{ \smallskip}
   &  He &  N &  O & Ne &  S & Ar & N/O  &   He &  N &  O & Ne &  S & Ar & N/O  \\
  \hline \noalign{ \smallskip}
   $\Delta$ & -0.04 &   0.35 &  -0.01 &  -0.10 &   0.53 &   0.00 &   0.36 &    -0.06 &   0.49 &  -0.10 &  -0.23 &   0.67 &   0.00 &   0.58 \\
   Number   & ( 36) &  ( 36) &  ( 37) &  ( 36) &  ( 32) &  (  0) &  ( 36) &    (  6) &  (  6) &  (  6) &  (  6) &  (  4) &  (  0) &  (  6) \\
   $\sigma$ &  0.08 &   0.32 &   0.18 &   0.23  &  0.61 &   0.00 &   0.31 &     0.06 &   0.30 &   0.14 &   0.23  &  0.41 &   0.00 &   0.24 \\
  \hline \noalign{ \smallskip}
  \end{tabular}
  \normalsize
 \end{table*}
  \begin{table*}[htb]
  \setlength{ \tabcolsep}{0.6mm}
  \caption{Same as Table~\ref{CompMeat} but for Monk et al.~(\cite{monk}).
}
  \label{CompMonk}
  \begin{tabular}{|l|l|l|l|l|l|l|l||l|l|l|l|l|l|l|}
  \noalign{ \smallskip}
  \hline \noalign{ \smallskip}
  & \multicolumn{7}{|c|}{ \bf \object{LMC}} & \multicolumn{7}{|c|}{ \bf \object{SMC}} \\
  \hline \noalign{ \smallskip}
  \hline \noalign{ \smallskip}
  \noalign{ \smallskip}
   &  He &  N &  O & Ne &  S & Ar & N/O  &   He &  N &  O & Ne &  S & Ar & N/O  \\
  \hline \noalign{ \smallskip}
   $\Delta$ &  0.00 &  -0.18 &  -0.11 &   0.01 &   0.00 &   0.00 &  -0.06 &    0.03 &   0.01 &  -0.15 &  -0.12 &   0.00 &   0.00 &   0.17 \\
   Number   & (  6) &  (  6) &  (  6) &  (  5) &  (  0) &  (  0) &  (  6) &   (  4) &  (  3) &  (  4) &  (  4) &  (  0) &  (  0) &  (  3) \\
   $\sigma$ &  0.05 &   0.14 &   0.17 &   0.03  &  0.00 &   0.00 &   0.18 &    0.06 &   0.06 &   0.08 &   0.12  &  0.00 &   0.00 &   0.10 \\
  \hline \noalign{ \smallskip}
  \end{tabular}
  \normalsize
 \end{table*}

When comparing abundances  derived  from  various
observations of the same object, but all derived with our method, 
we find small differences of about
$\pm$ 0.05-0.10 for helium, carbon, nitrogen, and oxygen. In some cases, 
especially in the \object{SMC}, 
 the difference for nitrogen can be about 4-5 times larger, which is probably 
due largely to  the use of a high nitrogen temperature, following the set of 
$T_{[NII]} = T_{[OIII]}$.  This question will be analyzed in more detail  
in a forthcoming paper where we intend to compare results from  detailed 
photo-ionization models with those of ICF's for the 
objects with the highest S/N spectra.
For neon and argon, the differences are $\pm$ 0.10-0.20 and reach $\pm$ 0.50  
for sulfur.\\
We summarize in Tables~\ref{CompMeat} and~\ref{CompMonk}  
the differences between the abundance determinations for various elements in
the Magellanic Clouds. 
The differences represent mainly the uncertainties in the observations, but
one should note that the uncertainties on intensities published in the 
literature appear to be larger in general than those of our own observations.

If we now compare published abundances and our own determinations with 
the same intensities as input, we sometimes find large differences. 
This reflects the uncertainties due to, mainly, atomic data, the ICF,
or the photo-ionization models. 
The agreement with MBC is usually excellent, as expected because we 
both use the same technique (ICF corrections). \\
The differences with MD are larger, reflecting the 
discrepancies between modeling and ICF methods.
We therefore find the usual differences here between ICF and
model determinations. \\
The abundance differences of oxygen or neon are acceptable (0.10-0.15),
but they are definitively too large for helium or nitrogen.
MD always finds a greater He abundance by about 0.05~dex (Table~\ref{CompMeat}). 
This difference becomes smaller (about 0.05~dex)
if we use only  the $He\,I\:\lambda_{5876}$ line. 
Uncertainty on the faint $\lambda_{4471}$ and $\lambda_{6678}$
He\,I lines could be one reason, but inaccuracies in the 
instrumental response curves is another factor that is not easy to estimate. 
For a few individual objects, large abundance differences found 
between various authors result from a markedly different temperature 
and/or density determination. Such cases will be discussed 
individually later.

\noindent
 We can summarize our own abundance determination uncertainties as follows~:
\begin{itemize}
  \item 0.05~dex or less for He
  \item about 0.10~dex for C, O and Ne
  \item about 0.15-0.20~dex for N ($~$because of larger ICF) and Ar
  \item up to $\sim$ 0.4-0.5 for S
\end{itemize}
Ratios, such as the N/O one, are of course much
less sensitive to temperatures or ICF errors and have been used whenever 
possible in the following.
\section{Results}
In the following, we discuss the various abundances patterns in detail. 
\subsection{Tables}

The physical parameters, c($H_{\beta}$), $n_{e^{-}}$, $T_{[OIII]}$,
$T_{[NII]}$, excitation class,  and total abundances for all the 
studied objects are presented in Tables~\ref{ABONDsmc} and~\ref{ABONDlmc}.
Mean abundances in the ISM of both Clouds, for reference, are taken
from H\,{\sc ii}~regions, as compiled by Dennefeld~(\cite{denn}). 
Another compilation by Garnett et al.~(\cite{Garn})
gives values in excellent  agreement  (within 0.05 dex).\\
The mean values for H\,{\sc ii}~regions are shown in the figures.
Solar values are from Grevesse~\&~Anders~(\cite{grev}).
Note that adopting the recent, lower, solar value proposed by 
Allende~Prieto~(\cite{alle1}) instead would not affect our 
discussion, as all comparisons are made with respect to the 
H\,{\sc ii}~regions values. 
Objects intermediate between "normal" PNe and "classical" Type$\,$I's
in the standard definition, which we called Type$\,$i in Paper~I, and defined 
as those objects having an excess in N/O but not in He.
 
 \begin{table*}[htb]
 \setlength{\tabcolsep}{0.6mm}
  \centering
  \caption{{\bf PNe Abundances in the SMC}.
The first column gives the identification of the object followed by an indication of the origin of the spectroscopic data: 3 letters indicate 
previous spectroscopic observations (mon for Monk et al.~\cite{monk}, mea for Meatheringham and Dopita~\cite{mea1};~\cite{mea2}, and vas 
for Vassiliadis et al.~\cite{vasi1}). 
Objects observed by us have no letters, except a  P1 mark for those  already discussed in Paper~I.
Columns 3 and 4 give the density and the temperature;  a $*$ symbol indicates that  no direct temperature or density determination
was possible so the value is chosen arbitrarily.
Columns 6 to 12 give the derived abundances with a quality symbol: poorly determined values are followed by ($:$); if one or more ionic 
stage is missing, we only have a lower limit to the abundance, hence the ($+$) symbol; and if no line at all was observed for a given
element, we use the faintest line intensity  detected in the spectrum to derive an upper limit to the abundance, hence the ($<$) symbol.}
  \label{ABONDsmc}
  \scriptsize
  \begin{tabular}{|lr|r|r|rr|r||r|r|r|r|r|r|r|r|r|r|r|r|}
  \noalign{\smallskip}
  \hline\noalign{\smallskip}
Name &  & Copt & $n_{e}$ & $T_{[OIII]}$ &  $T_{[NII]}$ & Exci & He &  C &  N &  O & Ne &  S & Ar &  C/N  &  C/O  &  N/O  &  C+N  & C+N+O \\
  \noalign{\smallskip}
  \hline\noalign{\smallskip}
  1 \object{SMP\_SMC\_1}          mea &   &  0.02 & *     5000 &  11100 & :11100 &  1.1 & $ $10.83 & $ $ 0.00 & $ $ 7.16 & $ $ 7.86 & $ $ 6.42 & $<$ 6.94 & $:$ 5.71  &  0.00 &  0.00 & -0.70 &  7.16 &  7.94 \\
  2 \object{SMP\_SMC\_2}          mea & I &  0.01 &       3000 &  16200 & :16200 &  6.3 & $ $11.10 & $ $ 0.00 & $ $ 7.47 & $ $ 8.01 & $ $ 7.21 & $ $ 7.32 & $ $ 5.78  &  0.00 &  0.00 & -0.54 &  7.47 &  8.12 \\
  3 \object{SMP\_SMC\_3}     (P1)     &   &  0.19 &      21500 &  13600 & :13600 &  2.7 & $ $10.94 & $ $ 8.77 & $ $ 7.00 & $ $ 7.98 & $ $ 7.00 & $ $ 6.76 & $:$ 5.49  &  1.76 &  0.79 & -0.98 &  8.78 &  8.84 \\
  4 \object{SMP\_SMC\_4}     (P1)     & I &  0.39 &       6100 &  15700 & :15700 &  3.1 & $ $11.04 & $<$ 6.55 & $:$ 7.33 & $ $ 7.83 & $ $ 6.89 & $+$ 7.34 & $ $ 5.44  & -0.78 & -1.28 & -0.50 &  7.40 &  7.97 \\
  5 \object{SMP\_SMC\_5}          mea & I &  0.00 &       3200 &  14500 & :14500 &  6.6 & $ $11.11 & $ $ 0.00 & $:$ 7.73 & $:$ 8.24 & $:$ 7.43 & $:$ 7.68 & $:$ 6.01  &  0.00 &  0.00 & -0.51 &  7.73 &  8.36 \\
  6 \object{SMP\_SMC\_6}     (P1)     & i &  0.64 &      11400 &  14400 &  15600 &  3.5 & $ $10.99 & $:$ 8.77 & $:$ 8.06 & $ $ 7.99 & $ $ 7.14 & $ $ 7.37 & $ $ 5.70  &  0.70 &  0.78 &  0.08 &  8.85 &  8.90 \\
  7 \object{SMP\_SMC\_7}          mea &   &  0.03 &       1900 &  17900 & :17900 &  4.2 & $ $10.83 & $ $ 0.00 & $ $ 7.27 & $ $ 7.96 & $ $ 7.18 & $ $ 7.27 & $ $ 5.51  &  0.00 &  0.00 & -0.69 &  7.27 &  8.04 \\
  8 \object{SMP\_SMC\_8}          mon &   &  0.26 & *     5000 &  13200 & :13200 &  2.8 & $ $10.91 & $ $ 0.00 & $ $ 0.00 & $ $ 7.98 & $ $ 6.96 & $ $ 0.00 & $ $ 5.07  &  0.00 &  0.00 &  0.00 &  0.00 &  0.00 \\
  9 \object{SMP\_SMC\_9}     (P1)     &   &  0.00 &       2200 &  13700 & ~13700 &  7.7 & $~$10.86 & $ $ 8.79 & $:$ 7.56 & $ $ 8.67 & $ $ 8.05 & $+$ 6.44 & $ $ 5.90  &  1.23 &  0.12 & -1.11 &  8.82 &  9.05 \\
 10 \object{SMP\_SMC\_10}    (P1)     &   &  0.26 &       1500 &  11300 &  11300 &  2.7 & $ $10.96 & $:$ 8.92 & $ $ 7.43 & $ $ 8.20 & $ $ 7.32 & $ $ 7.01 & $ $ 5.68  &  1.49 &  0.73 & -0.76 &  8.94 &  9.01 \\
 11 \object{SMP\_SMC\_11}             &   &  0.82 &        800 & *12000 & *12000 &  1.6 & $~$10.97 & $ $ 0.00 & $ $ 6.76 & $ $ 8.01 & $<$ 7.00 & $ $ 6.77 & $:$ 5.86  &  0.00 &  0.00 & -1.25 &  6.76 &  8.04 \\
 12 \object{SMP\_SMC\_12}             &   &  0.35 & *     5000 &  13200 & :13200 &  2.6 & $ $11.01 & $ $ 0.00 & $:$ 7.16 & $+$ 7.95 & $ $ 6.89 & $<$ 7.82 & $:$ 5.10  &  0.00 &  0.00 & -0.79 &  7.16 &  8.02 \\
 13 \object{SMP\_SMC\_13}    (P1)     & i &  0.16 &      10700 &  12400 & ~12400 &  3.6 & $ $10.95 & $:$ 8.64 & $:$ 7.58 & $ $ 8.17 & $ $ 7.10 & $ $ 7.19 & $ $ 5.62  &  1.06 &  0.47 & -0.59 &  8.68 &  8.79 \\
 14 \object{SMP\_SMC\_14}    (P1)     &   &  0.29 &       1900 &  13700 & ~13700 &  6.5 & $ $10.92 & $ $ 9.16 & $:$ 7.51 & $ $ 8.31 & $ $ 7.40 & $ $ 7.34 & $ $ 5.63  &  1.65 &  0.85 & -0.80 &  9.17 &  9.22 \\
 15 \object{SMP\_SMC\_15}         mea & I &  0.04 & *     5000 &  12000 &  16200 &  0.0 & $ $11.03 & $ $ 0.00 & $:$ 7.71 & $:$ 8.07 & $:$ 7.32 & $+$ 7.67 & $:$ 5.72  &  0.00 &  0.00 & -0.37 &  7.71 &  8.23 \\
 16 \object{SMP\_SMC\_16}         mea &   &  0.00 & *     5000 &  11800 & :11800 &  0.8 & $ $10.69 & $ $ 0.00 & $ $ 6.55 & $ $ 7.85 & $ $ 6.37 & $<$ 6.39 & $ $ 5.46  &  0.00 &  0.00 & -1.30 &  6.55 &  7.87 \\
 17 \object{SMP\_SMC\_17}         vas & I &  0.00 & *     5000 &  14800 & :14800 &  0.0 & $ $11.08 & $ $ 0.00 & $:$ 7.70 & $:$ 7.98 & $:$ 7.25 & $+$ 5.85 & $:$ 5.67  &  0.00 &  0.00 & -0.28 &  7.70 &  8.17 \\
 18 \object{SMP\_SMC\_18}         mon &   &  0.00 & *     5000 &  10700 & :10700 &  1.3 & $ $10.90 & $ $ 0.00 & $ $ 6.84 & $ $ 8.10 & $ $ 7.37 & $ $ 0.00 & $ $ 5.86  &  0.00 &  0.00 & -1.26 &  6.84 &  8.13 \\
 19 \object{SMP\_SMC\_19}    (P1)     & i &  0.21 &       2300 & *12000 & *12000 &  6.6 & $ $10.94 & $ $ 8.97 & $:$ 7.91 & $ $ 8.42 & $ $ 7.50 & $+$ 6.23 & $ $ 5.77  &  1.06 &  0.54 & -0.52 &  9.00 &  9.10 \\
 20 \object{SMP\_SMC\_20}    (P1)     & i &  0.05 & *     5000 &  12900 & :12900 &  1.9 & $ $10.95 & $:$ 8.36 & $ $ 7.65 & $ $ 7.83 & $ $ 6.74 & $<$ 7.56 & $ $ 5.21  &  0.71 &  0.53 & -0.18 &  8.44 &  8.53 \\
 21 \object{SMP\_SMC\_21}    (P1)     & i &  0.39 &      10200 &  24400 &  15700 &  6.5 & $~$11.00 & $<$ 7.04 & $:$ 7.91 & $ $ 7.56 & $ $ 6.82 & $ $ 7.05 & $ $ 5.58  & -0.86 & -0.51 &  0.35 &  7.96 &  8.11 \\
 22 \object{SMP\_SMC\_22}    (P1)     & I &  0.43 &       1700 &  24700 &  12200 &  7.9 & $ $11.06 & $ $ 7.23 & $:$ 7.99 & $ $ 7.59 & $ $ 6.75 & $ $ 6.60 & $ $ 5.47  & -0.76 & -0.36 &  0.40 &  8.06 &  8.18 \\
 23 \object{SMP\_SMC\_23}    (P1)     &   &  0.09 & *     5000 &  12800 & :12800 &  3.6 & $ $10.95 & $ $ 8.39 & $:$ 7.27 & $+$ 8.12 & $ $ 7.28 & $<$ 7.32 & $ $ 5.56  &  1.12 &  0.26 & -0.86 &  8.42 &  8.60 \\
 24 \object{SMP\_SMC\_24}         mon &   &  0.00 & *     5000 &  12700 & :12700 &  2.1 & $ $11.00 & $ $ 0.00 & $ $ 6.86 & $ $ 8.00 & $ $ 7.06 & $ $ 5.49 & $ $ 0.00  &  0.00 &  0.00 & -1.14 &  6.86 &  8.03 \\
 25 \object{SMP\_SMC\_25}    (P1)     & I &  0.23 &       9800 & :21100 &  21100 &  7.4 & $ $11.02 & $<$ 6.74 & $ $ 7.92 & $ $ 7.56 & $ $ 6.96 & $+$ 7.05 & $ $ 5.48  & -1.18 & -0.82 &  0.36 &  7.95 &  8.09 \\
 26 \object{SMP\_SMC\_26}             & i &  0.44 &        300 &  18000 &  11000 &  7.5 & $~$10.96 & $ $ 0.00 & $ $ 8.10 & $ $ 8.14 & $ $ 7.43 & $+$ 6.04 & $ $ 5.55  &  0.00 &  0.00 & -0.04 &  8.10 &  8.42 \\
 27 \object{SMP\_SMC\_27}         mon &   &  0.00 & *     5000 &  13100 & :13100 &  2.8 & $ $10.92 & $ $ 0.00 & $ $ 7.03 & $ $ 7.99 & $ $ 7.00 & $ $ 0.00 & $ $ 0.00  &  0.00 &  0.00 & -0.96 &  7.03 &  8.04 \\
 28 \object{SMP\_SMC\_28}    (P1)     & I &  0.19 &      20700 &  19200 &  14600 &  7.3 & $ $11.11 & $<$ 6.39 & $ $ 8.00 & $+$ 7.46 & $ $ 6.86 & $ $ 6.87 & $ $ 5.67  & -1.61 & -1.07 &  0.53 &  8.01 &  8.12 \\
 29 \object{SMP\_SMC\_32}             &   &  0.00 & *     5000 &  14700 & :14700 &  3.1 & $ $10.91 & $ $ 0.00 & $<$ 7.05 & $+$ 7.95 & $ $ 7.00 & $+$ 6.50 & $:$ 4.83  &  0.00 &  0.00 & -0.91 &  7.05 &  8.01 \\
 30 \object{SMP\_SMC\_34}             &   &  0.03 &        500 &  13000 &  10000 &  2.0 & $ $10.85 & $ $ 0.00 & $ $ 7.16 & $ $ 8.21 & $ $ 7.57 & $+$ 6.00 & $:$ 5.38  &  0.00 &  0.00 & -1.05 &  7.16 &  8.25 \\
 31 \object{MGPN\_SMC\_2}         vas &   &  1.31 & *     5000 & *12000 & *12000 &  2.4 & $<$10.52 & $ $ 0.00 & $<$ 6.67 & $ $ 8.19 & $ $ 7.42 & $<$ 7.67 & $:$ 6.00  &  0.00 &  0.00 & -1.52 &  6.67 &  8.20 \\
 32 \object{MGPN\_SMC\_7}             &   &  0.98 &        200 &  15500 & :15500 &  5.4 & $ $11.06 & $ $ 0.00 & $ $ 7.17 & $ $ 8.08 & $ $ 7.69 & $ $ 5.73 & $ $ 5.34  &  0.00 &  0.00 & -0.91 &  7.17 &  8.13 \\
 33 \object{MGPN\_SMC\_8}         vas &   &  0.00 & *     5000 & *12000 & *12000 &  5.9 & $ $10.74 & $ $ 0.00 & $ $ 7.45 & $ $ 8.44 & $<$ 7.45 & $+$ 6.15 & $<$ 5.78  &  0.00 &  0.00 & -1.00 &  7.45 &  8.48 \\
 34 \object{MGPN\_SMC\_9}             &   &  1.08 &       4600 &  20900 & :20900 &  1.8 & $ $10.81 & $ $ 0.00 & $ $ 6.49 & $ $ 7.35 & $ $ 6.49 & $ $ 6.72 & $:$ 5.27  &  0.00 &  0.00 & -0.86 &  6.49 &  7.41 \\
 35 \object{MGPN\_SMC\_12}   (P1)     & I &  0.56 &       1300 &  22100 &  12800 &  7.5 & $ $11.21 & $:$ 8.40 & $ $ 8.14 & $ $ 7.47 & $ $ 7.10 & $ $ 6.67 & $ $ 5.82  &  0.26 &  0.93 &  0.67 &  8.59 &  8.63 \\
 36 \object{MGPN\_SMC\_13}            &   &  0.28 &       1000 &  18800 & :18800 & 10.0 & $~$11.04 & $ $ 0.00 & $:$ 7.68 & $ $ 8.49 & $ $ 7.57 & $~$ 8.31 & $ $ 5.52  &  0.00 &  0.00 & -0.81 &  7.68 &  8.55 \\
 37 \object{LHA\_115-N\_8}        mon &   &  0.00 & *     5000 & *12000 & *12000 &  0.0 & $ $ 0.00 & $ $ 0.00 & $ $ 6.74 & $ $ 8.20 & $:$ 0.00 & $ $ 6.15 & $ $ 0.00  &  0.00 &  0.00 & -1.46 &  6.74 &  8.21 \\
  \noalign{\smallskip}
  \hline\noalign{\smallskip}
\multicolumn{7}{|l|}{\bf Mean HII (Dennefeld 1989)} & 10.90 &  7.19 &  6.46 &  7.97 &  7.22 &  6.32 &  5.78 &  0.73 & -0.78 & -1.51 &  7.26 &  8.05 \\
\multicolumn{7}{|l|}{\bf Solar values (Grevesse and Anders 1989)} & 10.99 &  8.56 &  8.05 &  8.93 &  8.09 &  7.21 &  6.56 &  0.51 & -0.37 & -0.88 &  8.68
&  9.12 \\
  \hline\noalign{\smallskip}
  \noalign{\bigskip}
  \hline\noalign{\smallskip}
\multicolumn{7}{|l|}{\bf Type I} & 11.09 &  7.82 &  7.78 &  7.80 &  7.09 &  7.01 &  5.67 & -0.25 &  0.29 & -0.03 &  7.85 &  8.21 \\
\multicolumn{7}{|l|}{\it Number} & (  9) & (  2) & (  9) & (  9) & (  9) & (  9) & (  9) & (  2) & (  2) & (  9) & (  9) & (  9) \\
\multicolumn{7}{|l|}{$\sigma$} &  0.06 &  0.59 &  0.25 &  0.27 &  0.22 &  0.55 &  0.18 &  0.51 &  0.64 &  0.47 &  0.34 &  0.18 \\
  \noalign{\smallskip}
  \hline\noalign{\smallskip}
\multicolumn{7}{|l|}{\bf Type I+i} & 11.04 &  8.39 &  7.81 &  7.89 &  7.10 &  6.92 &  5.63 &  0.51 &  0.48 & -0.08 &  8.11 &  8.38 \\
\multicolumn{7}{|l|}{\it Number} & ( 15) & (  6) & ( 15) & ( 15) & ( 15) & ( 14) & ( 15) & (  6) & (  6) & ( 15) & ( 15) & ( 15) \\
\multicolumn{7}{|l|}{$\sigma$} &  0.07 &  0.56 &  0.23 &  0.29 &  0.25 &  0.56 &  0.18 &  0.62 &  0.41 &  0.43 &  0.48 &  0.33 \\
  \noalign{\smallskip}
  \hline\noalign{\smallskip}
\multicolumn{7}{|l|}{\bf non-Type I} & 10.90 &  8.81 &  7.11 &  8.09 &  7.16 &  6.62 &  5.51 &  1.45 &  0.55 & -0.98 &  7.44 &  8.29 \\
\multicolumn{7}{|l|}{\it Number} & ( 20) & (  5) & ( 19) & ( 22) & ( 19) & ( 14) & ( 18) & (  5) & (  5) & ( 19) & ( 21) & ( 21) \\
\multicolumn{7}{|l|}{$\sigma$} &  0.09 &  0.25 &  0.33 &  0.26 &  0.43 &  0.70 &  0.30 &  0.24 &  0.30 &  0.21 &  0.83 &  0.44 \\
  \noalign{\smallskip}
  \hline\noalign{\smallskip}
\multicolumn{19}{l}{\bf Differences $<$PNe$>$ - $<$HII$>$} \\
  \noalign{\smallskip}
  \hline\noalign{\smallskip}
\multicolumn{7}{|l|}{\bf Type I} &  0.19 &  0.63 &  1.32 & -0.17 & -0.13 &  0.69 & -0.11 & -0.98 &  1.07 &  1.48 &  0.58 &  0.16 \\
\multicolumn{7}{|l|}{\bf Type I+i} &  0.14 &  1.20 &  1.35 & -0.08 & -0.12 &  0.60 & -0.15 & -0.22 &  1.26 &  1.43 &  0.84 &  0.33 \\
\multicolumn{7}{|l|}{\bf non-Type I} &  0.00 &  1.62 &  0.65 &  0.12 & -0.06 &  0.30 & -0.27 &  0.72 &  1.33 &  0.53 &  0.17 &  0.24 \\
  \noalign{\smallskip}
  \hline\noalign{\smallskip}
  \end{tabular}
  \normalsize
 \setlength{\tabcolsep}{6pt}
 \end{table*}

 \begin{table*}[htb]
 \setlength{\tabcolsep}{0.6mm}
  \centering
  \caption{{\bf PNe Abundances in the LMC}. Symbols and 
headings as in SMC table.}
  \label{ABONDlmc}
  \scriptsize
  \begin{tabular}{|lr|r|r|rr|r||r|r|r|r|r|r|r|r|r|r|r|r|}
  \noalign{\smallskip}
  \hline\noalign{\smallskip}
Name &  & Copt & $n_{e}$ & $T_{[OIII]}$ &  $T_{[NII]}$ & Exci & He &  C &  N &  O & Ne &  S & Ar &  C/N  &  C/O  &  N/O  &  C+N  & C+N+O \\
  \noalign{\smallskip}
  \hline\noalign{\smallskip}
  1 \object{SMP\_LMC\_1}      (P1)      & i &  0.38 &       2600 &  10800 &   9400 &  3.5 & $ $10.95 & $:$ 8.40 & $:$ 7.91 & $ $ 8.34 & $ $ 7.44 & $ $ 7.32 & $ $ 5.94  &  0.50 &  0.06 & -0.43 &  8.52 &  8.74 \\
  2 \object{SMP\_LMC\_2}            mea &   &  0.01 & *     5000 &  12100 & :12100 &  1.1 & $~$10.96 & $ $ 0.00 & $ $ 6.95 & $ $ 8.03 & $<$ 6.25 & $<$ 6.62 & $:$ 6.00  &  0.00 &  0.00 & -1.09 &  6.95 &  8.07 \\
  3 \object{SMP\_LMC\_3}            mea &   &  0.06 &       4000 &  14000 & :14000 &  1.8 & $ $10.96 & $ $ 0.00 & $ $ 7.03 & $ $ 7.75 & $ $ 6.82 & $ $ 6.76 & $ $ 5.35  &  0.00 &  0.00 & -0.71 &  7.03 &  7.82 \\
  4 \object{SMP\_LMC\_4}                &   &  0.32 & *     5000 &  11800 & :11800 &  6.7 & $ $11.07 & $ $ 0.00 & $ $ 7.93 & $ $ 8.61 & $ $ 7.78 & $<$ 7.87 & $:$ 5.91  &  0.00 &  0.00 & -0.68 &  7.93 &  8.69 \\
  5 \object{SMP\_LMC\_5}            mea &   &  0.09 & *     5000 &  13200 & :13200 &  1.2 & $ $10.79 & $ $ 0.00 & $ $ 6.34 & $ $ 8.00 & $ $ 6.64 & $<$ 6.44 & $:$ 5.35  &  0.00 &  0.00 & -1.66 &  6.34 &  8.01 \\
  6 \object{SMP\_LMC\_6}            mea &   &  0.04 &      11800 &  13300 & :13300 &  6.1 & $ $10.97 & $ $ 0.00 & $:$ 7.52 & $:$ 8.51 & $:$ 7.75 & $:$ 7.09 & $:$ 5.99  &  0.00 &  0.00 & -0.99 &  7.52 &  8.55 \\
  7 \object{SMP\_LMC\_7}            mea & I &  0.00 &       1000 &  19600 & :19600 &  8.8 & $~$11.15 & $ $ 0.00 & $:$ 8.58 & $:$ 8.33 & $:$ 7.44 & $:$ 7.41 & $:$ 6.08  &  0.00 &  0.00 &  0.26 &  8.58 &  8.77 \\
  8 \object{SMP\_LMC\_8}            mea & I &  0.01 & *     5000 &  11100 & :11100 &  2.5 & $ $11.14 & $ $ 0.00 & $ $ 8.00 & $ $ 8.16 & $ $ 7.26 & $+$ 8.04 & $:$ 6.22  &  0.00 &  0.00 & -0.16 &  8.00 &  8.38 \\
  9 \object{SMP\_LMC\_9}            vas & i &  0.25 &       3000 &  14800 & :14800 &  6.6 & $ $10.97 & $ $ 0.00 & $:$ 7.82 & $:$ 8.31 & $:$ 7.74 & $+$ 6.63 & $:$ 6.04  &  0.00 &  0.00 & -0.48 &  7.82 &  8.43 \\
 10 \object{SMP\_LMC\_10}           vas & I &  0.26 &       2800 &  17900 & :17900 &  0.0 & $ $11.00 & $ $ 0.00 & $:$ 7.55 & $:$ 7.82 & $:$ 7.21 & $+$ 6.00 & $:$ 5.81  &  0.00 &  0.00 & -0.27 &  7.55 &  8.01 \\
 11 \object{SMP\_LMC\_11}               & i &  1.06 &       6200 &  29100 &  20800 &  5.0 & $ $10.78 & $ $ 0.00 & $ $ 7.10 & $ $ 7.18 & $ $ 6.74 & $ $ 6.06 & $:$ 4.99  &  0.00 &  0.00 & -0.08 &  7.10 &  7.44 \\
 12 \object{SMP\_LMC\_13}     (P1)      &   &  0.00 & *     5000 &  13000 & :13000 &  6.3 & $ $11.03 & $ $ 7.92 & $:$ 7.19 & $ $ 8.39 & $ $ 7.60 & $<$ 7.85 & $ $ 5.87  &  0.72 & -0.47 & -1.19 &  7.99 &  8.53 \\
 13 \object{SMP\_LMC\_14}           mea & I &  0.02 &        300 &  21000 & :21000 & 10.0 & $~$11.32 & $ $ 0.00 & $:$ 8.52 & $:$ 8.15 & $:$ 7.72 & $:$ 7.52 & $:$ 6.31  &  0.00 &  0.00 &  0.37 &  8.52 &  8.67 \\
 14 \object{SMP\_LMC\_15}           mea &   &  0.08 &       3800 &  14000 & :14000 &  6.1 & $ $11.03 & $ $ 0.00 & $:$ 7.47 & $:$ 8.26 & $:$ 7.57 & $+$ 6.41 & $:$ 5.97  &  0.00 &  0.00 & -0.79 &  7.47 &  8.32 \\
 15 \object{SMP\_LMC\_16}               & I &  0.68 &        600 &  19800 &  12000 &  8.2 & $ $11.00 & $ $ 0.00 & $ $ 8.65 & $ $ 8.32 & $ $ 7.38 & $ $ 7.29 & $ $ 6.00  &  0.00 &  0.00 &  0.33 &  8.65 &  8.81 \\
 16 \object{SMP\_LMC\_17}               & I &  0.85 &        400 & :13500 &  13500 &  6.0 & $~$11.02 & $ $ 0.00 & $ $ 8.62 & $ $ 8.25 & $ $ 7.92 & $ $ 6.88 & $ $ 6.20  &  0.00 &  0.00 &  0.38 &  8.62 &  8.77 \\
 17 \object{SMP\_LMC\_18}           vas & I &  0.20 & *     5000 &  16600 & :16600 &  0.0 & $~$11.08 & $ $ 0.00 & $<$ 7.49 & $+$ 7.87 & $ $ 7.00 & $<$ 8.30 & $<$ 5.56  &  0.00 &  0.00 & -0.38 &  7.49 &  8.02 \\
 18 \object{SMP\_LMC\_19}           mea &   &  0.04 &       2000 &  13900 & :13900 &  6.9 & $ $11.08 & $ $ 0.00 & $:$ 7.82 & $:$ 8.50 & $:$ 7.77 & $+$ 6.62 & $:$ 6.18  &  0.00 &  0.00 & -0.68 &  7.82 &  8.58 \\
 19 \object{SMP\_LMC\_20}           mea & I &  0.02 &       2200 &  19300 &  13400 &  8.2 & $~$11.06 & $ $ 0.00 & $:$ 8.39 & $:$ 8.12 & $:$ 7.59 & $:$ 7.24 & $:$ 6.22  &  0.00 &  0.00 &  0.27 &  8.39 &  8.58 \\
 20 \object{SMP\_LMC\_21}           vas & I &  0.16 &       2800 &  25200 & :25200 &  7.4 & $ $11.11 & $ $ 0.00 & $ $ 8.37 & $ $ 7.86 & $ $ 7.19 & $ $ 7.21 & $ $ 5.98  &  0.00 &  0.00 &  0.51 &  8.37 &  8.48 \\
 21 \object{SMP\_LMC\_23}           mea &   &  0.02 & *     5000 &  11300 & :11300 &  0.0 & $ $10.91 & $ $ 0.00 & $:$ 7.10 & $:$ 8.30 & $:$ 7.31 & $<$ 7.51 & $:$ 5.98  &  0.00 &  0.00 & -1.20 &  7.10 &  8.33 \\
 22 \object{SMP\_LMC\_24}           mea &   &  0.01 &        800 &  12800 & :12800 &  8.8 & $+$11.15 & $ $ 0.00 & $ $ 7.96 & $ $ 8.61 & $ $ 8.03 & $+$ 6.59 & $:$ 6.36  &  0.00 &  0.00 & -0.66 &  7.96 &  8.70 \\
 23 \object{SMP\_LMC\_25}               &   &  0.51 &      16600 &  13000 &  10900 &  3.5 & $ $10.90 & $ $ 0.00 & $ $ 7.26 & $ $ 8.17 & $ $ 7.28 & $ $ 6.73 & $ $ 5.71  &  0.00 &  0.00 & -0.91 &  7.26 &  8.22 \\
 24 \object{SMP\_LMC\_26}           mon & i &  0.00 & *     5000 & :13600 &  13600 &  0.0 & $ $ 9.85 & $ $ 0.00 & $ $ 6.93 & $ $ 7.06 & $:$ 6.90 & $ $ 0.00 & $ $ 0.00  &  0.00 &  0.00 & -0.13 &  6.93 &  7.30 \\
 25 \object{SMP\_LMC\_27}               &   &  0.12 & *     5000 &  11100 & :11100 &  3.5 & $ $10.98 & $ $ 0.00 & $ $ 7.10 & $ $ 8.31 & $ $ 7.42 & $+$ 7.68 & $:$ 6.06  &  0.00 &  0.00 & -1.20 &  7.10 &  8.33 \\
 26 \object{SMP\_LMC\_29}           mea & I &  0.01 &       4100 &  19800 & :19800 &  8.2 & $ $11.15 & $ $ 0.00 & $ $ 8.79 & $ $ 8.05 & $ $ 7.22 & $ $ 7.88 & $ $ 6.27  &  0.00 &  0.00 &  0.74 &  8.79 &  8.86 \\
 27 \object{SMP\_LMC\_30}               & I &  0.39 &        400 &  15700 &  12300 &  8.0 & $ $11.15 & $ $ 0.00 & $ $ 8.55 & $ $ 8.24 & $ $ 7.78 & $ $ 7.29 & $ $ 6.19  &  0.00 &  0.00 &  0.32 &  8.55 &  8.72 \\
 28 \object{SMP\_LMC\_31}           mea & i &  0.79 &       6600 &  14000 &  16100 &  0.4 & $ $10.79 & $ $ 0.00 & $ $ 7.03 & $ $ 7.29 & $ $ 5.58 & $ $ 6.27 & $:$ 5.56  &  0.00 &  0.00 & -0.26 &  7.03 &  7.48 \\
 29 \object{SMP\_LMC\_32}           mea &   &  0.09 &       3900 &  16100 & :16100 &  8.3 & $ $11.08 & $ $ 0.00 & $ $ 7.73 & $ $ 8.39 & $ $ 7.56 & $ $ 7.41 & $ $ 6.05  &  0.00 &  0.00 & -0.65 &  7.73 &  8.47 \\
 30 \object{SMP\_LMC\_33}               &   &  0.63 &       5600 &  13400 &   7800 &  6.5 & $ $10.90 & $ $ 0.00 & $ $ 7.84 & $ $ 8.91 & $ $ 8.21 & $ $ 6.83 & $ $ 5.90  &  0.00 &  0.00 & -1.07 &  7.84 &  8.95 \\
 31 \object{SMP\_LMC\_35}           mea & i &  0.04 &       1500 &  13500 & :13500 &  5.3 & $ $10.97 & $ $ 0.00 & $:$ 7.88 & $:$ 8.37 & $:$ 7.48 & $:$ 7.38 & $:$ 6.08  &  0.00 &  0.00 & -0.49 &  7.88 &  8.49 \\
 32 \object{SMP\_LMC\_36}           mon &   &  0.41 & *     5000 &  15500 & :15500 &  6.5 & $ $11.00 & $ $ 0.00 & $ $ 7.72 & $ $ 8.36 & $ $ 7.72 & $ $ 0.00 & $ $ 5.64  &  0.00 &  0.00 & -0.64 &  7.72 &  8.45 \\
 33 \object{SMP\_LMC\_37}           mea & i &  0.03 &       5900 &  14300 &  11000 &  6.8 & $ $10.93 & $ $ 0.00 & $:$ 8.27 & $:$ 8.52 & $:$ 7.62 & $:$ 7.36 & $:$ 6.00  &  0.00 &  0.00 & -0.25 &  8.27 &  8.72 \\
 34 \object{SMP\_LMC\_38}           mea & I &  0.02 &       8800 &  13100 & :13100 &  0.0 & $ $11.10 & $ $ 0.00 & $:$ 7.73 & $:$ 8.27 & $:$ 7.55 & $:$ 7.37 & $:$ 6.25  &  0.00 &  0.00 & -0.54 &  7.73 &  8.38 \\
 35 \object{SMP\_LMC\_40}               & I &  0.16 &        800 &  13500 & ~13500 &  7.5 & $ $11.02 & $ $ 0.00 & $ $ 8.14 & $ $ 8.56 & $ $ 7.84 & $ $ 7.56 & $ $ 5.98  &  0.00 &  0.00 & -0.42 &  8.14 &  8.70 \\
 36 \object{SMP\_LMC\_41}           mea & I &  0.01 &       1100 &  15400 & :15400 &  7.6 & $ $11.15 & $ $ 0.00 & $:$ 8.39 & $:$ 8.38 & $:$ 7.63 & $:$ 7.54 & $:$ 6.17  &  0.00 &  0.00 &  0.01 &  8.39 &  8.69 \\
 37 \object{SMP\_LMC\_42}           mea & I &  0.05 &       2500 &  15100 & :15100 &  0.0 & $~$11.10 & $ $ 0.00 & $:$ 7.88 & $:$ 8.03 & $:$ 7.12 & $<$ 8.15 & $:$ 5.80  &  0.00 &  0.00 & -0.15 &  7.88 &  8.26 \\
 38 \object{SMP\_LMC\_44}               & I &  0.13 &        900 &  16700 &  11300 &  7.5 & $ $11.04 & $ $ 0.00 & $ $ 8.30 & $ $ 8.47 & $ $ 8.01 & $ $ 6.89 & $ $ 6.08  &  0.00 &  0.00 & -0.17 &  8.30 &  8.69 \\
 39 \object{SMP\_LMC\_45}           mea & I &  0.14 &       1000 &  16300 & :16300 &  5.4 & $ $11.00 & $ $ 0.00 & $:$ 7.66 & $:$ 8.05 & $:$ 7.35 & $+$ 6.46 & $:$ 5.94  &  0.00 &  0.00 & -0.39 &  7.66 &  8.19 \\
 40 \object{SMP\_LMC\_46}           mea & i &  0.01 &       3900 &  14900 & :14900 &  6.1 & $ $10.91 & $ $ 0.00 & $:$ 8.07 & $:$ 8.28 & $:$ 7.54 & $+$ 6.78 & $:$ 5.98  &  0.00 &  0.00 & -0.21 &  8.07 &  8.49 \\
 41 \object{SMP\_LMC\_47}     (P1)      & I &  0.42 &       4800 &  14700 &  13400 &  6.6 & $ $11.09 & $ $ 8.56 & $ $ 8.69 & $ $ 8.25 & $ $ 7.54 & $ $ 7.50 & $ $ 6.13  & -0.14 &  0.31 &  0.45 &  8.93 &  9.01 \\
 42 \object{SMP\_LMC\_48}               & i &  0.25 &       1900 &  11300 &   8800 &  3.0 & $ $10.92 & $ $ 0.00 & $:$ 7.71 & $ $ 8.24 & $ $ 7.14 & $ $ 7.18 & $:$ 5.88  &  0.00 &  0.00 & -0.53 &  7.71 &  8.35 \\
 43 \object{SMP\_LMC\_49}               &   &  0.04 &       1000 &  13300 &  21600 &  6.0 & $ $11.05 & $ $ 0.00 & $ $ 7.60 & $ $ 8.33 & $ $ 7.48 & $ $ 7.63 & $ $ 6.05  &  0.00 &  0.00 & -0.72 &  7.60 &  8.40 \\
 44 \object{SMP\_LMC\_50}           mea &   &  0.00 &       3600 &  13700 & :13700 &  5.4 & $ $11.01 & $ $ 0.00 & $:$ 7.42 & $:$ 8.22 & $:$ 7.35 & $~$ 7.59 & $:$ 5.84  &  0.00 &  0.00 & -0.80 &  7.42 &  8.28 \\
 45 \object{SMP\_LMC\_51}           mon &   &  0.00 & *     5000 &  12000 & :12000 &  4.5 & $ $11.01 & $ $ 0.00 & $ $ 7.48 & $ $ 8.32 & $ $ 7.25 & $ $ 6.20 & $ $ 0.00  &  0.00 &  0.00 & -0.84 &  7.48 &  8.38 \\
 46 \object{SMP\_LMC\_52}           mea &   &  0.02 &       2900 &  12000 & *12000 &  5.8 & $ $10.92 & $ $ 0.00 & $:$ 7.59 & $:$ 8.55 & $:$ 7.72 & $:$ 7.73 & $:$ 6.16  &  0.00 &  0.00 & -0.96 &  7.59 &  8.59 \\
 47 \object{SMP\_LMC\_53}     (P1)      & i &  0.37 &       4800 &  14100 &  13400 &  2.7 & $ $10.93 & $<$ 6.70 & $ $ 7.72 & $ $ 8.23 & $ $ 7.61 & $ $ 7.26 & $ $ 5.86  & -1.02 & -1.53 & -0.51 &  7.76 &  8.35 \\
 48 \object{SMP\_LMC\_54}     (P1)      & I &  0.11 &        400 & ~11700 &  11700 &  9.3 & $~$11.10 & $<$ 8.14 & $:$ 8.84 & $:$ 8.74 & $:$ 8.08 & $:$ 7.84 & $:$ 6.63  & -0.70 & -0.59 &  0.11 &  8.92 &  9.14 \\
 49 \object{SMP\_LMC\_55}           mea &   &  0.02 & *    47400 &  12400 &  10100 &  0.7 & $ $10.82 & $ $ 0.00 & $ $ 7.27 & $ $ 8.40 & $ $ 6.80 & $ $ 6.44 & $:$ 5.76  &  0.00 &  0.00 & -1.14 &  7.27 &  8.44 \\
 50 \object{SMP\_LMC\_56}           mea &   &  0.00 & *     5000 &  13100 & :13100 &  0.0 & $<$10.79 & $ $ 0.00 & $:$ 6.57 & $:$ 7.93 & $:$ 6.51 & $<$ 6.63 & $:$ 5.78  &  0.00 &  0.00 & -1.36 &  6.57 &  7.95 \\
 51 \object{SMP\_LMC\_58}           mea & i &  0.00 & *     5000 &  12600 & :12600 &  0.0 & $ $10.97 & $ $ 0.00 & $:$ 7.58 & $:$ 8.10 & $:$ 7.09 & $+$ 7.98 & $:$ 5.83  &  0.00 &  0.00 & -0.53 &  7.58 &  8.21 \\
 52 \object{SMP\_LMC\_59}               & I &  1.04 &        200 &  15000 &  12400 &  8.6 & $~$11.05 & $ $ 0.00 & $ $ 8.49 & $ $ 8.52 & $ $ 8.09 & $ $ 7.12 & $:$ 6.20  &  0.00 &  0.00 & -0.03 &  8.49 &  8.81 \\
 53 \object{SMP\_LMC\_60}           mea &   &  0.00 & *     5000 &  16300 & :16300 &  9.9 & $ $11.00 & $ $ 0.00 & $ $ 7.97 & $ $ 8.87 & $ $ 8.05 & $<$ 8.15 & $ $ 6.05  &  0.00 &  0.00 & -0.90 &  7.97 &  8.92 \\
 54 \object{SMP\_LMC\_61}           mea &   &  0.03 &      36200 &  11100 &  10200 &  0.0 & $ $11.08 & $ $ 0.00 & $:$ 7.32 & $:$ 8.53 & $:$ 7.76 & $:$ 6.76 & $:$ 6.26  &  0.00 &  0.00 & -1.22 &  7.32 &  8.56 \\
 55 \object{SMP\_LMC\_62}           mea & I &  0.06 &       3400 &  15900 & :15900 &  5.9 & $ $11.01 & $ $ 0.00 & $:$ 7.89 & $:$ 8.15 & $:$ 7.34 & $+$ 6.66 & $:$ 5.73  &  0.00 &  0.00 & -0.26 &  7.89 &  8.34 \\
 56 \object{SMP\_LMC\_63}     (P1)      &   &  0.26 &       2600 &  11100 &  10000 &  4.1 & $ $10.92 & $:$ 8.80 & $:$ 7.72 & $ $ 8.39 & $ $ 7.53 & $ $ 7.28 & $ $ 5.94  &  1.08 &  0.41 & -0.66 &  8.84 &  8.97 \\
 57 \object{SMP\_LMC\_64}           mea &   &  0.03 & *     5000 & *12000 & *12000 &  0.1 & $ $10.76 & $ $ 0.00 & $ $ 6.34 & $ $ 7.05 & $<$ 5.99 & $+$ 7.09 & $:$ 5.66  &  0.00 &  0.00 & -0.71 &  6.34 &  7.12 \\
 58 \object{SMP\_LMC\_65}           mea &   &  0.01 & *     5000 &  10700 & :10700 &  0.0 & $+$11.05 & $ $ 0.00 & $<$ 7.28 & $:$ 8.35 & $:$ 7.34 & $<$ 8.29 & $:$ 5.97  &  0.00 &  0.00 & -1.08 &  7.28 &  8.39 \\
 59 \object{SMP\_LMC\_66}     (P1)      &   &  0.11 &       3300 &  13000 & :13000 &  5.5 & $ $10.97 & $ $ 8.51 & $ $ 7.61 & $ $ 8.31 & $ $ 7.29 & $ $ 7.93 & $ $ 5.91  &  0.89 &  0.19 & -0.70 &  8.56 &  8.76 \\
 60 \object{SMP\_LMC\_67}     (P1)      & i &  0.06 &       8100 &  10000 &   8600 &  1.4 & $ $10.92 & $<$ 7.66 & $ $ 8.00 & $ $ 8.50 & $ $ 7.60 & $ $ 6.46 & $:$ 6.04  & -0.34 & -0.84 & -0.50 &  8.16 &  8.66 \\
 61 \object{SMP\_LMC\_68}               &   &  0.00 &       1900 &  12000 & *12000 &  9.9 & $~$11.02 & $ $ 0.00 & $:$ 7.88 & $ $ 8.94 & $ $ 8.15 & $ $ 8.76 & $ $ 6.34  &  0.00 &  0.00 & -1.06 &  7.88 &  8.98 \\
 62 \object{SMP\_LMC\_69}     (P1)      & I &  0.29 &        200 &  17200 &  12700 &  9.0 & $~$11.06 & $<$ 8.32 & $ $ 8.61 & $ $ 8.63 & $ $ 8.10 & $+$ 6.85 & $:$ 5.95  & -0.29 & -0.31 & -0.02 &  8.79 &  9.01 \\
 63 \object{SMP\_LMC\_71}           mon &   &  0.15 & *     5000 &  13000 &  10800 &  6.7 & $ $11.04 & $ $ 0.00 & $ $ 8.03 & $ $ 8.63 & $ $ 7.85 & $ $ 6.61 & $ $ 0.00  &  0.00 &  0.00 & -0.60 &  8.03 &  8.73 \\
 64 \object{SMP\_LMC\_72}               &   &  0.21 &        100 &  16000 & :16000 &  9.9 & $~$11.02 & $ $ 0.00 & $:$ 7.88 & $:$ 8.92 & $:$ 8.08 & $:$ 8.23 & $:$ 6.10  &  0.00 &  0.00 & -1.04 &  7.88 &  8.96 \\
 65 \object{SMP\_LMC\_73}     (P1)      &   &  0.34 &       4500 &  11700 &   9700 &  5.7 & $ $10.94 & $ $ 8.78 & $:$ 8.03 & $ $ 8.66 & $ $ 7.80 & $ $ 7.37 & $ $ 6.10  &  0.75 &  0.12 & -0.63 &  8.85 &  9.07 \\
 66 \object{SMP\_LMC\_74}           mea &   &  0.03 &       4900 &  12600 & :12600 &  5.8 & $ $11.03 & $ $ 0.00 & $:$ 7.65 & $:$ 8.42 & $:$ 7.65 & $:$ 7.26 & $:$ 6.07  &  0.00 &  0.00 & -0.77 &  7.65 &  8.49 \\
 67 \object{SMP\_LMC\_75}           mon &   &  0.25 & *     5000 &  12100 & :12100 &  4.2 & $ $10.96 & $ $ 0.00 & $ $ 7.44 & $ $ 8.29 & $ $ 7.56 & $ $ 0.00 & $ $ 0.00  &  0.00 &  0.00 & -0.85 &  7.44 &  8.34 \\
 68 \object{SMP\_LMC\_76}           mea &   &  0.07 & *     5000 &  11500 & :11500 &  0.0 & $~$11.00 & $ $ 0.00 & $:$ 7.33 & $:$ 8.15 & $:$ 7.27 & $<$ 7.30 & $:$ 5.78  &  0.00 &  0.00 & -0.82 &  7.33 &  8.21 \\
 69 \object{SMP\_LMC\_77}     (P1)      &   &  0.94 &       4700 &  13300 &  10400 &  1.7 & $ $10.86 & $:$ 9.12 & $:$ 6.73 & $ $ 8.27 & $ $ 7.48 & $ $ 5.89 & $:$ 5.46  &  2.39 &  0.85 & -1.54 &  9.12 &  9.18 \\
 70 \object{SMP\_LMC\_78}           mea & I &  0.06 &       4200 &  14200 & :14200 &  6.0 & $ $11.01 & $ $ 0.00 & $:$ 7.88 & $:$ 8.39 & $:$ 7.56 & $:$ 7.49 & $:$ 6.04  &  0.00 &  0.00 & -0.51 &  7.88 &  8.51 \\
 71 \object{SMP\_LMC\_79}               & I &  0.28 &       3100 &  12900 &  13600 &  5.3 & $ $11.03 & $ $ 0.00 & $ $ 8.02 & $ $ 8.34 & $ $ 7.47 & $ $ 7.67 & $ $ 5.93  &  0.00 &  0.00 & -0.32 &  8.02 &  8.51 \\
 72 \object{SMP\_LMC\_80}               &   &  0.08 &      36500 &  10600 & ~10600 &  2.4 & $ $10.93 & $ $ 0.00 & $ $ 7.39 & $ $ 8.34 & $ $ 7.32 & $ $ 7.12 & $:$ 6.13  &  0.00 &  0.00 & -0.95 &  7.39 &  8.38 \\
 73 \object{SMP\_LMC\_81}           mon &   &  0.15 &      11300 &  14300 & :14300 &  5.1 & $ $10.89 & $ $ 0.00 & $ $ 7.12 & $ $ 8.25 & $ $ 7.32 & $ $ 6.21 & $ $ 0.00  &  0.00 &  0.00 & -1.12 &  7.12 &  8.28 \\
 74 \object{SMP\_LMC\_82}               & I &  0.47 &       4900 &  16100 &  16200 &  7.6 & $ $11.25 & $ $ 0.00 & $ $ 8.59 & $ $ 8.16 & $ $ 7.56 & $ $ 7.72 & $ $ 6.21  &  0.00 &  0.00 &  0.43 &  8.59 &  8.73 \\
 75 \object{SMP\_LMC\_83}           mea & I &  0.00 &       2000 &  17800 &  12000 &  7.8 & $ $11.08 & $ $ 0.00 & $:$ 8.13 & $:$ 8.23 & $:$ 7.45 & $:$ 7.42 & $:$ 6.24  &  0.00 &  0.00 & -0.11 &  8.13 &  8.48 \\
 76 \object{SMP\_LMC\_84}               &   &  0.20 &       2900 &  11900 & ~11900 &  3.0 & $ $10.97 & $ $ 0.00 & $ $ 7.43 & $ $ 8.15 & $ $ 7.17 & $ $ 7.30 & $:$ 5.79  &  0.00 &  0.00 & -0.72 &  7.43 &  8.23 \\
 77 \object{SMP\_LMC\_85}               &   &  0.42 & ~    31400 &  11700 &  13900 &  1.6 & $ $10.93 & $ $ 0.00 & $ $ 7.26 & $ $ 8.00 & $ $ 6.83 & $ $ 6.67 & $ $ 5.73  &  0.00 &  0.00 & -0.74 &  7.26 &  8.08 \\
 78 \object{SMP\_LMC\_86}               & I &  0.59 &       1200 &  17400 &  12500 &  7.9 & $ $11.19 & $ $ 0.00 & $ $ 8.66 & $ $ 8.01 & $ $ 7.40 & $ $ 7.27 & $:$ 5.99  &  0.00 &  0.00 &  0.65 &  8.66 &  8.75 \\
 79 \object{SMP\_LMC\_87}               & I &  0.24 &       2000 &  18700 &  12100 &  8.6 & $ $11.17 & $ $ 0.00 & $ $ 8.74 & $ $ 8.23 & $ $ 7.49 & $ $ 7.26 & $ $ 6.11  &  0.00 &  0.00 &  0.50 &  8.74 &  8.86 \\
 80 \object{SMP\_LMC\_88}     (P1)      & I &  1.31 &       3400 &  19900 &  18000 &  9.2 & $~$11.11 & $<$ 8.84 & $ $ 7.79 & $ $ 7.91 & $ $ 7.51 & $ $ 6.77 & $ $ 5.61  &  1.04 &  0.93 & -0.11 &  8.88 &  8.92 \\
  \noalign{\smallskip}
  \hline\noalign{\smallskip}
  \end{tabular}
  \normalsize
 \setlength{\tabcolsep}{6pt}
 \end{table*}
 \newpage
 \clearpage
 \begin{table*}[htb]
 \setlength{\tabcolsep}{0.6mm}
  \addtocounter{table}{-1}
  \centering
  \caption{{\bf PNe Abundances in the LMC} (continued)}
  \scriptsize
  \begin{tabular}{|lr|r|r|rr|r||r|r|r|r|r|r|r|r|r|r|r|r|}
  \noalign{\smallskip}
  \hline\noalign{\smallskip}
Name &  & Copt & $n_{e}$ & $T_{[OIII]}$ &  $T_{[NII]}$ & Exci & He &  C &  N &  O & Ne &  S & Ar &  C/N  &  C/O  &  N/O  &  C+N  & C+N+O \\
  \noalign{\smallskip}
  \hline\noalign{\smallskip}
 81 \object{SMP\_LMC\_89}           mea &   &  0.00 &       5500 &  13500 & :13500 &  5.0 & $ $10.94 & $ $ 0.00 & $:$ 7.52 & $:$ 8.37 & $:$ 7.64 & $+$ 6.23 & $:$ 5.81  &  0.00 &  0.00 & -0.84 &  7.52 &  8.43 \\
 82 \object{SMP\_LMC\_90}               & i &  0.52 &        700 &  19200 &   9400 &  5.7 & $ $10.91 & $ $ 0.00 & $ $ 7.84 & $ $ 8.18 & $ $ 7.51 & $ $ 6.13 & $:$ 5.63  &  0.00 &  0.00 & -0.35 &  7.84 &  8.35 \\
 83 \object{SMP\_LMC\_91}               & I &  0.46 &        700 &  17900 &  12300 &  8.9 & $ $11.18 & $ $ 0.00 & $ $ 8.48 & $ $ 8.27 & $ $ 7.77 & $ $ 7.16 & $ $ 6.02  &  0.00 &  0.00 &  0.21 &  8.48 &  8.69 \\
 84 \object{SMP\_LMC\_92}    (P1)       &   &  0.31 &       5300 &  13000 &  10000 &  6.1 & $ $10.92 & $ $ 8.21 & $ $ 7.77 & $ $ 8.62 & $ $ 7.83 & $ $ 7.20 & $:$ 6.16  &  0.44 & -0.41 & -0.85 &  8.34 &  8.81 \\
 85 \object{SMP\_LMC\_93}               & i &  0.57 &        300 &  17100 &  11000 &  7.6 & $ $10.96 & $ $ 0.00 & $ $ 8.69 & $ $ 8.58 & $ $ 8.10 & $ $ 7.31 & $:$ 6.06  &  0.00 &  0.00 &  0.11 &  8.69 &  8.94 \\
 86 \object{SMP\_LMC\_94}           mea &   &  0.93 & *     5000 & *12000 & *12000 &  7.8 & $ $11.08 & $ $ 0.00 & $<$ 5.98 & $ $ 7.29 & $<$ 6.48 & $:$ 6.09 & $ $ 5.81  &  0.00 &  0.00 & -1.31 &  5.98 &  7.31 \\
 87 \object{SMP\_LMC\_95}           vas & i &  0.45 &       1000 &  13700 & :13700 &  6.0 & $+$10.96 & $ $ 0.00 & $:$ 7.95 & $:$ 8.41 & $:$ 7.76 & $+$ 6.51 & $:$ 6.26  &  0.00 &  0.00 & -0.46 &  7.95 &  8.54 \\
 88 \object{SMP\_LMC\_96}           mea & I &  0.00 &       1200 &  23100 &  12100 &  9.6 & $~$11.25 & $ $ 0.00 & $ $ 8.32 & $ $ 8.04 & $ $ 7.52 & $ $ 7.09 & $ $ 6.15  &  0.00 &  0.00 &  0.28 &  8.32 &  8.51 \\
 89 \object{SMP\_LMC\_97}           mea &   &  0.04 &       1300 &  14900 & :14900 &  8.1 & $~$10.98 & $ $ 0.00 & $:$ 7.78 & $ $ 8.50 & $ $ 7.54 & $+$ 8.01 & $ $ 5.97  &  0.00 &  0.00 & -0.72 &  7.78 &  8.57 \\
 90 \object{SMP\_LMC\_98}               &   &  0.42 &       7000 &  12300 &  10100 &  5.3 & $ $10.93 & $ $ 0.00 & $:$ 7.79 & $ $ 8.57 & $ $ 7.77 & $ $ 7.22 & $:$ 6.05  &  0.00 &  0.00 & -0.78 &  7.79 &  8.63 \\
 91 \object{SMP\_LMC\_99}    (P1)       & i &  0.24 &       3600 &  12200 &  13200 &  5.7 & $ $10.92 & $ $ 8.69 & $ $ 8.15 & $ $ 8.56 & $ $ 7.62 & $ $ 7.65 & $ $ 6.62  &  0.55 &  0.14 & -0.41 &  8.80 &  9.00 \\
 92 \object{SMP\_LMC\_100}   (P1)       &   &  0.38 &       3700 &  14000 & :14000 &  6.4 & $ $11.02 & $ $ 8.55 & $ $ 7.61 & $ $ 8.36 & $ $ 7.56 & $ $ 7.49 & $ $ 5.97  &  0.94 &  0.19 & -0.75 &  8.60 &  8.80 \\
 93 \object{SMP\_LMC\_101}          mea &   &  0.04 &       4200 &  15300 & :15300 &  8.5 & $ $11.00 & $ $ 0.00 & $:$ 7.65 & $:$ 8.50 & $:$ 7.85 & $+$ 6.66 & $:$ 5.88  &  0.00 &  0.00 & -0.86 &  7.65 &  8.56 \\
 94 \object{SMP\_LMC\_102}          vas &   &  0.07 & *     5000 &  17000 & :17000 &  8.7 & $ $11.00 & $ $ 0.00 & $<$ 7.29 & $ $ 8.28 & $ $ 7.36 & $<$ 8.09 & $:$ 5.88  &  0.00 &  0.00 & -0.99 &  7.29 &  8.32 \\
 95 \object{SMP\_LMC\_104}          vas &   &  0.26 & *     5000 &  16900 & :16900 &  9.3 & $~$11.10 & $ $ 0.00 & $<$ 7.22 & $ $ 8.56 & $ $ 7.71 & $<$ 8.03 & $ $ 5.99  &  0.00 &  0.00 & -1.35 &  7.22 &  8.58 \\
 96 \object{SMP\_LMC\_104a}  (P1)       & i &  0.45 &       3700 & :15900 &  15900 &  7.8 & $ $10.92 & $ $ 8.56 & $:$ 7.51 & $ $ 7.78 & $ $ 7.19 & $ $ 7.65 & $ $ 5.36  &  1.05 &  0.79 & -0.26 &  8.60 &  8.66 \\
 97 \object{SMP\_LMC\_122}              & i &  0.34 &        500 &  15300 & ~15300 &  9.3 & $:$10.93 & $ $ 0.00 & $ $ 8.83 & $ $ 8.94 & $ $ 8.08 & $ $ 8.09 & $:$ 5.92  &  0.00 &  0.00 & -0.10 &  8.83 &  9.19 \\
 98 \object{LMC-PNazzo}                 &   &  0.89 &       1000 & *12000 & *12000 &  2.5 & $~$11.02 & $ $ 0.00 & $ $ 7.79 & $ $ 8.59 & $ $ 8.15 & $ $ 7.06 & $:$ 5.96  &  0.00 &  0.00 & -0.80 &  7.79 &  8.66 \\
 99 \object{MGPN\_LMC\_3}           vas &   &  0.25 &        500 &  13500 & :13500 &  2.6 & $<$10.77 & $ $ 0.00 & $ $ 7.50 & $ $ 8.23 & $ $ 7.80 & $+$ 6.08 & $:$ 6.06  &  0.00 &  0.00 & -0.74 &  7.50 &  8.31 \\
100 \object{MGPN\_LMC\_7}           vas &   &  0.00 &       7900 &  13300 & :13300 &  5.1 & $+$10.94 & $ $ 0.00 & $ $ 7.69 & $ $ 8.43 & $ $ 7.78 & $+$ 6.67 & $<$ 6.11  &  0.00 &  0.00 & -0.75 &  7.69 &  8.50 \\
101 \object{MGPN\_LMC\_21}          vas & I &  0.00 & *     5000 &  17100 & :17100 & 10.0 & $+$11.15 & $ $ 0.00 & $<$ 7.80 & $+$ 8.05 & $ $ 7.12 & $<$ 8.60 & $<$ 5.95  &  0.00 &  0.00 & -0.24 &  7.80 &  8.24 \\
102 \object{MGPN\_LMC\_29}          vas & I &  0.18 &        800 &  15200 & :15200 &  8.2 & $+$11.12 & $ $ 0.00 & $:$ 7.99 & $:$ 8.33 & $:$ 7.60 & $+$ 6.70 & $:$ 6.33  &  0.00 &  0.00 & -0.34 &  7.99 &  8.50 \\
103 \object{MGPN\_LMC\_35}          vas & i &  0.11 & *     5000 & *12000 & *12000 &  7.9 & $ $10.81 & $ $ 0.00 & $:$ 8.03 & $ $ 8.53 & $ $ 8.08 & $<$ 7.78 & $<$ 5.62  &  0.00 &  0.00 & -0.50 &  8.03 &  8.65 \\
104 \object{MGPN\_LMC\_39}          vas & i &  0.49 & *     5000 & *12000 & *12000 &  0.5 & $ $10.72 & $ $ 0.00 & $ $ 7.09 & $ $ 7.56 & $<$ 6.60 & $<$ 6.93 & $:$ 5.85  &  0.00 &  0.00 & -0.48 &  7.09 &  7.69 \\
105 \object{MGPN\_LMC\_44}          vas & I &  0.31 & *     5000 &  16300 & :16300 &  8.8 & $+$11.15 & $ $ 0.00 & $<$ 7.68 & $+$ 8.13 & $ $ 7.22 & $<$ 8.50 & $<$ 6.16  &  0.00 &  0.00 & -0.45 &  7.68 &  8.26 \\
106 \object{MGPN\_LMC\_45}          vas & I &  0.86 & *     5000 &  15100 & :15100 &  5.3 & $ $11.04 & $ $ 0.00 & $ $ 8.41 & $ $ 8.27 & $ $ 7.50 & $<$ 7.91 & $ $ 6.01  &  0.00 &  0.00 &  0.14 &  8.41 &  8.65 \\
107 \object{MGPN\_LMC\_46}          vas & I &  0.24 &        100 &  17700 & :17700 &  9.1 & $+$11.13 & $ $ 0.00 & $:$ 8.46 & $:$ 8.10 & $:$ 7.56 & $+$ 6.70 & $:$ 6.30  &  0.00 &  0.00 &  0.36 &  8.46 &  8.62 \\
108 \object{MGPN\_LMC\_74}          vas & I &  0.00 &        600 & *12000 & *12000 &  0.9 & $ $11.06 & $ $ 0.00 & $ $ 7.46 & $ $ 7.85 & $ $ 6.80 & $+$ 6.02 & $:$ 5.97  &  0.00 &  0.00 & -0.39 &  7.46 &  8.00 \\
109 \object{MGPN\_LMC\_83}          vas & I &  0.53 & *     5000 & *12000 & *12000 & 10.0 & $+$11.45 & $ $ 0.00 & $ $ 8.76 & $ $ 9.17 & $ $ 8.52 & $<$ 8.00 & $:$ 6.93  &  0.00 &  0.00 & -0.42 &  8.76 &  9.31 \\
110 \object{[M94b] 5}               vas & I &  0.16 &        200 &  17200 & :17200 &  8.0 & $+$11.28 & $ $ 0.00 & $ $ 8.19 & $ $ 8.08 & $ $ 7.41 & $+$ 6.38 & $<$ 6.27  &  0.00 &  0.00 &  0.11 &  8.19 &  8.44 \\
111 \object{[M94b] 41}              vas &   &  0.68 & *     5000 &  16400 & :16400 &  2.1 & $~$11.07 & $ $ 0.00 & $ $ 7.25 & $ $ 8.04 & $<$ 6.63 & $+$ 5.87 & $:$ 5.73  &  0.00 &  0.00 & -0.79 &  7.25 &  8.10 \\
112 \object{[M94b] 42}              vas & I &  0.41 &        300 &  21000 & :21000 &  8.8 & $+$11.21 & $ $ 0.00 & $ $ 8.22 & $ $ 7.88 & $ $ 7.39 & $+$ 6.38 & $:$ 6.18  &  0.00 &  0.00 &  0.34 &  8.22 &  8.38 \\
113 \object{[M94b] 43}              vas & I &  0.14 &       1200 &  16800 & :16800 & 10.0 & $ $11.06 & $ $ 0.00 & $:$ 8.10 & $:$ 8.55 & $:$ 7.70 & $+$ 6.67 & $:$ 6.03  &  0.00 &  0.00 & -0.44 &  8.10 &  8.68 \\
114 \object{[M94b] 45}              vas & I &  0.63 &        100 &  18800 & :18800 &  8.5 & $+$11.09 & $ $ 0.00 & $:$ 8.18 & $:$ 8.13 & $:$ 7.74 & $+$ 6.45 & $:$ 6.06  &  0.00 &  0.00 &  0.05 &  8.18 &  8.45 \\
115 \object{[M94b] 46}              vas &   &  0.22 & *     5000 &  13500 & :13500 &  5.4 & $+$10.85 & $ $ 0.00 & $ $ 7.73 & $ $ 8.36 & $ $ 7.67 & $+$ 6.54 & $:$ 6.12  &  0.00 &  0.00 & -0.63 &  7.73 &  8.45 \\
116 \object{[M94b] 48}              vas &   &  0.05 &      10600 &  16800 & :16800 &  0.9 & $ $10.88 & $ $ 0.00 & $ $ 6.85 & $ $ 7.66 & $<$ 5.91 & $+$ 5.43 & $ $ 5.61  &  0.00 &  0.00 & -0.82 &  6.85 &  7.73 \\
117 \object{[M94b] 49}              vas & I &  0.14 &       2800 &  14700 & :14700 &  7.0 & $ $11.08 & $ $ 0.00 & $ $ 8.04 & $ $ 8.35 & $ $ 7.56 & $+$ 6.57 & $ $ 6.00  &  0.00 &  0.00 & -0.31 &  8.04 &  8.52 \\
118 \object{[M94b] 50}              vas &   &  0.00 & *     5000 & *12000 & *12000 &  0.0 & $+$10.76 & $ $ 0.00 & $:$ 7.64 & $:$ 8.48 & $:$ 7.75 & $+$ 6.73 & $:$ 6.23  &  0.00 &  0.00 & -0.83 &  7.64 &  8.54 \\
119 \object{[M94b] 52}              vas &   &  0.00 &        300 & *12000 & *12000 &  2.3 & $~$11.22 & $ $ 0.00 & $ $ 7.25 & $ $ 8.27 & $ $ 7.64 & $+$ 6.37 & $:$ 6.07  &  0.00 &  0.00 & -1.02 &  7.25 &  8.31 \\
120 \object{LHA 120-N 99}           mon &   &  0.25 & *     5000 & :11700 &  11700 &  0.0 & $ $10.04 & $ $ 0.00 & $ $ 7.18 & $ $ 8.51 & $:$ 6.87 & $ $ 0.00 & $ $ 0.00  &  0.00 &  0.00 & -1.32 &  7.18 &  8.53 \\
 \noalign{\smallskip}
  \hline\noalign{\smallskip}
\multicolumn{7}{|l|}{\bf Mean HII (Dennefeld 1989)} & 10.93 &  7.87 &  6.97 &  8.38 &  7.64 &  6.67 &  6.20 &  0.90 & -0.51 & -1.41 &  7.92 &  8.51 \\
\multicolumn{7}{|l|}{\bf Solar values (Grevesse and Anders 1989)} & 10.99 &  8.56 &  8.05 &  8.93 &  8.09 &  7.21 &  6.56 &  0.51 & -0.37 & -0.88 &  8.68
&  9.12 \\
  \hline\noalign{\smallskip}
  \noalign{\bigskip}
  \hline\noalign{\smallskip}
\multicolumn{7}{|l|}{\bf Type I} & 11.12 &  8.56 &  8.28 &  8.22 &  7.54 &  7.09 &  6.11 & -0.14 &  0.31 &  0.04 &  8.27 &  8.59 \\
\multicolumn{7}{|l|}{\it Number} & ( 43) & (  1) & ( 40) & ( 43) & ( 43) & ( 37) & ( 39) & (  1) & (  1) & ( 40) & ( 43) & ( 43) \\
\multicolumn{7}{|l|}{$\sigma$} &  0.09 &  0.00 &  0.36 &  0.26 &  0.32 &  0.51 &  0.23 &  0.00 &  0.00 &  0.35 &  0.40 &  0.29 \\
  \noalign{\smallskip}
  \hline\noalign{\smallskip}
\multicolumn{7}{|l|}{\bf Type I+i} & 11.03 &  8.55 &  8.12 &  8.20 &  7.50 &  7.08 &  6.04 &  0.49 &  0.33 & -0.09 &  8.16 &  8.52 \\
\multicolumn{7}{|l|}{\it Number} & ( 63) & (  4) & ( 60) & ( 63) & ( 62) & ( 54) & ( 57) & (  4) & (  4) & ( 60) & ( 63) & ( 63) \\
\multicolumn{7}{|l|}{$\sigma$} &  0.20 &  0.10 &  0.47 &  0.35 &  0.42 &  0.54 &  0.29 &  0.42 &  0.28 &  0.35 &  0.49 &  0.39 \\
  \noalign{\smallskip}
  \hline\noalign{\smallskip}
\multicolumn{7}{|l|}{\bf non-Type I} & 10.96 &  8.56 &  7.45 &  8.33 &  7.54 &  6.93 &  5.93 &  1.03 &  0.13 & -0.90 &  7.55 &  8.44 \\
\multicolumn{7}{|l|}{\it Number} & ( 55) & (  7) & ( 53) & ( 57) & ( 52) & ( 43) & ( 51) & (  7) & (  7) & ( 53) & ( 57) & ( 57) \\
\multicolumn{7}{|l|}{$\sigma$} &  0.15 &  0.37 &  0.40 &  0.34 &  0.38 &  0.68 &  0.22 &  0.59 &  0.42 &  0.24 &  0.58 &  0.38 \\
  \noalign{\smallskip}
  \hline\noalign{\smallskip}
\multicolumn{19}{l}{\bf Differences $<$PNe$>$ - $<$HII$>$} \\
  \noalign{\smallskip}
  \hline\noalign{\smallskip}
\multicolumn{7}{|l|}{\bf Type I} &  0.19 &  0.69 &  1.31 & -0.16 & -0.10 &  0.42 & -0.09 & -1.04 &  0.82 &  1.45 &  0.35 &  0.08 \\
\multicolumn{7}{|l|}{\bf Type I+i} &  0.10 &  0.68 &  1.15 & -0.18 & -0.14 &  0.41 & -0.16 & -0.41 &  0.84 &  1.32 &  0.24 &  0.01 \\
\multicolumn{7}{|l|}{\bf non-Type I} &  0.03 &  0.69 &  0.48 & -0.05 & -0.10 &  0.26 & -0.27 &  0.13 &  0.64 &  0.51 & -0.37 & -0.07 \\
  \noalign{\smallskip}
  \hline\noalign{\smallskip}
  \end{tabular}
  \normalsize
 \setlength{\tabcolsep}{6pt}
 \end{table*}

 \begin{table*}[htb]
 \setlength{\tabcolsep}{0.6mm}
  \centering
  \caption{Objects with peculiar emission lines}
  \label{AutresRaies}
  \begin{tabular}{|l|c|c|c|c|c|l|}
  \noalign{\smallskip}
  \hline\noalign{\smallskip}
Name & $N\,III_{4641}$ & $C\,III_{4649}$ & $C\,IV_{4659}$ & $C\,IV_{5806}$ & Iron  & Comments \\
  \noalign{\smallskip}
  \hline\noalign{\smallskip}
\multicolumn{7}{|c|}{\bf SMC} \\
  \hline\noalign{\smallskip}
SMP SMC 3	& 	& 3.1	& 	&       & 		& 	\\
SMP SMC 6	& 1.8	& 	& 1.9	& Yes   & 		& 	\\
SMP SMC 10	& 0.4	& 	& 	&       & 		& 	\\
SMP SMC 14	& 	& 1.5	& 	&       & 		& 	\\
SMP SMC 19	& 	& 1.3	& 	&       & 		& 	\\
SMP SMC 21	& 	& 	&$\sim$6.0 &    & [FeVII] 	& 	\\
SMP SMC 22	& 	& 	& 1.1	&       & [FeVII]	& 	\\
MGPN SMC 9	& 	& 	& 1.0	&       & 		& Wind ? ($H_\alpha$ profil) 	\\
MGPN SMC 12	& 	& 	& 2.8	&       & 		& 	\\
MGPN SMC 13	& 	& 	& 5.2	& Possible & 		& 	\\
  \noalign{\smallskip}		        		        
  \hline\noalign{\smallskip}	        		        
\multicolumn{7}{|c|}{\bf LMC} \\        		        
  \hline\noalign{\smallskip}	        		        
SMP LMC 4	& 4.4	& 	& 	&       & 		& 	\\
SMP LMC 13	& 	& 	&$\sim$1.3&     & 	        & 	\\
SMP LMC 17	& 	& 	& 	&       & [FeVII]	& 	\\
SMP LMC 44	& 	&$\sim$3.7& 	&       & 	        & $\lambda_{6856}$	\\
SMP LMC 47	& 3.4	& 	& 1.5	&       & 		& 	\\
SMP LMC 66	& 2.3	& 	& 	&       & 		& 	\\
SMP LMC 68	& 	& 4.3	&	& Yes   & 		& OIII $\lambda_{5590}$; $\lambda\lambda_{6050-6120}$; $\lambda_{6527}$	\\
SMP LMC 73	& 1.3	& 	& 	& Possible & 		& 	\\
SMP LMC 82	& 	& 	& 	&       & [FeVII]	& 	\\
SMP LMC 85	& 	& 1.2	& 	&       & 		& 	\\
SMP LMC 86	& 	& 	& 3.1	&       & 		& 	\\
SMP LMC 87	& 2.5	& 	& 	&       & 		& 	\\
SMP LMC 88	& 	& 	& 	&       & [FeVII]	& $\lambda_{6826}$	\\
SMP LMC 92	&$\sim$2.0& 	& 	&       & 	        & 	\\
SMP LMC 93	& 	& 	& 4.5	&       & 		& 	\\
SMP LMC 98	& 1.2	& 	& 	&       & 		& 	\\
SMP LMC 99	& 3.8	& 	& 	& Possible & 		& 	\\
SMP LMC 100	& 1.9	& 	& 1.4	& Possible & 		& OIII $\lambda_{3133}$	$\lambda_{3444}$\\
SMP LMC 104a	& 	& 	&$\sim$1.2 &    & [FeVI],[FeVII]& $\lambda_{5620}$; $\lambda_{6837}$	\\
  \noalign{\smallskip}
  \hline\noalign{\smallskip}
  \end{tabular}
  \normalsize
 \setlength{\tabcolsep}{6pt}
 \end{table*}

 \begin{table*}[htb]
 \setlength{\tabcolsep}{0.6mm}
  \caption{Extended objects}
  \label{Etendus}
  \begin{tabular}{|l|l|l|}
  \noalign{\smallskip}
  \hline\noalign{\smallskip}
Name & Size (arcsec) & Comments \\
  \noalign{\smallskip}
  \hline\noalign{\smallskip}
\multicolumn{3}{|c|}{\bf LMC} \\
  \hline\noalign{\smallskip}
SMP LMC 11	& $>$6-8''			& 2 blobs (extended lines, V $>$ 100km/s) 	\\
SMP LMC 17	& $\sim$3-4'' - $\sim$6-7''	& Bipolar + 2 blobs	\\
SMP LMC 24	& $\sim$1.5'' 			& 	\\
SMP LMC 27	& blob at $\sim$4''		& Thin Shell + Northern Blob 	\\
SMP LMC 60	& $\sim$1.5'' 			& 	\\
SMP LMC 86	& blob at $\sim$4-5''		& Southern extension	\\
SMP LMC 90	& $\sim$1.5''			& 	\\
SMP LMC 91	& $\sim$1.5'' 			& 	\\
SMP LMC 122	& $\sim$1.5''			& 	\\
Jacoby LMC 26	& $\sim$2-3''			& 	\\
  \noalign{\smallskip}
  \hline\noalign{\smallskip}
  \end{tabular}
  \normalsize
 \setlength{\tabcolsep}{6pt}
 \end{table*}

\clearpage

In Table~\ref{AutresRaies} 
we note the presence of some  peculiar emission lines, 
identified in  a few objects of our sample (and also 
seen in WR stars).
 The observed intensities are given in the same units as before
($H_{\beta}$=100).
The symbol $\sim$ indicates only a marginal detection.
In the  literature, only 5 objects are classified as Wolf Rayet objects~: 
\begin{itemize}
 \item \object{SMP SMC 6} and \object{MGPN SMC 8}
 \item \object{SMP LMC 38}, \object{SMP LMC 58} and \object{SMP LMC 61}
\end{itemize}
\noindent
We point out in Table~\ref{Etendus} some objects that we observed in
$H_{\alpha}$ imaging, and that appear to be resolved. 
Some comments  on their morphology  are given also. 

Although we should now discuss the various abundance patterns, but in any 
diagram, such as for instance the first one where we compare N/O to He/H 
(Fig.~\ref{nohe}) for the 183 objects, 
objects lying outside the average location of the PNe call attention to themselves.
Such "outliers" could be very interesting in having 
extremely high (or low) abundances compared to the average value, provided 
we are certain this abundance determination is not affected by larger-than-average
uncertainties (due to a poor quality spectrum, or a poor temperature 
determination, etc).
We therefore checked all those objects one by one. 
In some cases, the spectra are of poor quality with only a few bright lines 
available, and they account for the upper limits in He seen in Fig.~\ref{nohe} 
and will be removed from further analysis.
In total, 27 objects were eliminated due to poor quality spectra, eight 
 in the \object{SMC}~ and nineteen in the \object{LMC}.

Note also that for 8 additional objects with better spectra, 
the temperature and or density could not be derived directly from line ratios,
so that an arbitrary value had to be assigned. This is provisionally 
adequate as the range of densities and temperatures in PNe is usually limited, 
but these objects deserve special care in the discussion. There are  
5 in the \object{LMC}: \object{SMP LMC 64}, \object{SMP LMC 94}, 
\object{MGPN LMC 35}, \object{MGPN LMC 39}, and  \object{[M94b] 50}; 
and 3 in the \object{SMC}: \object{MGPN SMC 2}, \object{MGPN SMC 8},
and \object{LHA 115-N 8}.  

The removed objects are essentially from the data of  
Vassiliadis et al.~(\cite{vasi1}): all of them deserve
further observations, and will be studied again in the future.
The N/O versus He/H diagram is then redrawn in  Fig.~\ref{noheNEW} with 
the remaining 156 objects which will be the basis of further analysis.  
Remaining outliers (which cannot be ascribed 
a priori to a poor spectrum or a poor temperature/density determination) 
will be discussed when they appear in the following figures 
(Figs.~\ref{noheNEW}-\ref{nearNEW}), but  all are collected together in 
Table~\ref{objPROBLEM}. 
%
 \begin{table*}
  \setlength{\tabcolsep}{0.6mm}
  \centering
  \caption{Outstanding objects in various diagrams.
For each of them, we give in the first line its name, the adopted temperature and
density, and the elemental abundances that seem to be abnormal.
The second line gives the ICF used in the corresponding column, so that a check 
can be made of whether the problem comes from there or from another source.
A $*$ symbol in front of the density or the temperature indicates that this
value could not be derived directly from line ratios (see Sect.~\ref{secTempDens}).}
  \label{objPROBLEM}
  \begin{tabular}{|l|r|rr|c|c|c|c|c|c|c|c|c|}
   \noalign{\bigskip}
   \hline\noalign{\smallskip}
Name	& Ne-	& $T_{OIII}$ & $T_{NII}$& N/O-He	& He-Ar 	& N/O-N 	& N/O-O 	& N/O-Ar	& Ne-O  	& O-Ar      \\
   \hline\noalign{\smallskip}															    
   \hline\noalign{\smallskip}															    
SMP SMC 1	& *5000	& 11000	& 11000	& 		& 		& 		& 		&       	& 6.42,7.86	& 	    \\
		& 	&	&	& 		& 		& 		& 		&       	& 1.22,1.01	& 	    \\
SMP SMC 16	& *5000	& 11800	& 11800	& -1.30,10.69	& 		& 		& 		&       	& 6.37,7.85	& 	    \\
		& 	&	&	& 2.04/1.02	& 		& 		& 		&       	& 2.02,1.02	& 	    \\
SMP SMC 32	&  5000	& 14700	& 14700	&	 	& 		& 		& 		& -0.91,4.83   	& 		& 7.95,4.83 \\
		&  	& 	& 	&	 	& 		& 		& 		& 134/1.14-1.14	& 		& 1.14-1.73 \\
MGPN SMC 2		& *5000	&*12000	&*12000	& -1.52$<$10.52	& 10.52,6.00	& 		& 		&       	& 		& 	    \\
		& 	&	&	& 		& 1.73		& 		& 		&       	& 		& 	    \\
MGPN SMC 12	&  1300	& 22100	& 12800	&  0.67,11.21	& 		&  		&  0.67,7.47	&       	& 7.10,7.47	& 	    \\
		& 	&	&	& 5.24/1.62	& 		& 		& 	1.62	&       	& 2.34,1.62	& 	    \\
   \hline\noalign{\smallskip}						        								    
SMP LMC 11	&  6200	& 29100	& 20800	& -0.08,10.78	& 		& -0.08,7.10	& -0.08,7.18	& -0.08,4.99	& 		&	    \\
		&  	& 	&	& 3.07/1.30	& 		& 	3.07	& 	1.30	& 3.07/1.30-2.04& 		&	    \\
SMP LMC 26	& *5000	& 13600	& 13600	& -0.13,09.85	& 		& -0.13,6.93	& -0.13,7.06	&       	& 6.90,7.06	& 	    \\
		& 	& 	&	& 1.00/1.00	& 		& 	1.00	& 	1.00	&       	& 1.00-1.00	& 	    \\
SMP LMC 31	&  6600	& 14000	& 16100	& 		& 		& 		& 		&       	& 5.58,7.29	& 	    \\
		& 	& 	&	& 		& 		& 		& 		&       	& 1.92-1.02	& 	    \\
SMP LMC 54	&   400	& 11700	& 11700	& 		& 11.10,6.63	&   		& 		&  0.11.6.63	& 		& 	    \\
		& 	& 	&	& 		& 3.85		&   		& 		& 2.96/2.09-3.85& 		& 	    \\
SMP LMC 55	& 47400	& 12200	&  9600	& 		& 		& 		& 		&       	& 6.80,8.40	& 	    \\
		& 	& 	&	& 		& 		& 		& 		&       	& 9.00-1.01	& 	    \\
SMP LMC 56	& *5000	& 13100	& 13100	& -1.36,10.79	& 		& 		& 		&       	& 		& 	    \\
		& 	& 	&	& 2.48/1.14	& 		& 		& 		&       	& 		& 	    \\
SMP LMC 64$^{\dagger}$& *5000	&*12000	&*12000	& -0.71,10.76	& 	& 		& -0.71,7.05	& 	   	& 		& 7.05,5.66 \\
		& 	&	&	& 1.80/1.01	& 		& 		& 	1.01	& 	   	& 		& 1.01-1.73 \\
SMP LMC 99	&  3600	& 12200	& 12200	& 		& 10.92,6.62	& 		& 		& -0.41,6.62	& 		& 	    \\
		& 	& 	&	& 		& 1.43		& 		& 		& 20.2/1.32-1.43& 		& 	    \\
MGPN LMC 39	& *5000	& 25100	& 25100	& -0.48,10.72	& 		& 	 	& 		&       	& 		& 	    \\
		& 	& 	&	& 2.60/1.07	& 	 	& 		&       	& 		& 		&	    \\
$[M94b]$ 8	& 10600	& 16800	& 16800	& 		& 		& 		& 		&       	& $<$5.91,7.66	& 	    \\
		& 	& 	&	& 		& 		& 		& 		&       	& 2.71-1.03	& 	    \\
   \hline\noalign{\smallskip}
  \end{tabular}
  \normalsize
\\
\begin{flushleft}
$^{\dagger}$ Dopita and Meatheringham~\cite{dopSMP64} have dedicated a paper to study 
of this peculiar high density object.
\end{flushleft}
 \end{table*}

\subsection{Helium and nitrogen}
As already discussed in our first paper, the Type$\,$I PNe were introduced 
by Peimbert~(\cite{peim1},\cite{peim2}) for Galactic objects, because they 
could be clearly separated from other PNe in an N/O versus He/H diagram. 
We present such a diagram in Fig.~\ref{noheNEW}, now with a large sample
according to the original definition, but corrected for the different 
metallicity in the \object{LMC} or \object{SMC}, as discussed in Paper~I.  
There is a smooth continuity between the two types,so that it now becomes difficult
to separate the PNe in these 2 different classes from observational data 
alone (although the distinction is clear on theoretical grounds). 
Because of the addition of many faint nebulae, the dispersion of the points
has also increased.
Type$\,$i objects, introduced in Paper~I, appear in the upper left part of 
the diagram (following their definition:  high N/O but not high He).
Here also, the separation between Type$\,$I and Type$\,$i is now   
arbitrary, as both classes, with high N/O, cover a continuum in He abundances.
Clearly the original abundances (at the time 
of the formation of the progenitor) need to be known in order to distinguish 
the various objects by their level of enrichment. 

\begin{figure}[htbp]
   \includegraphics[angle=-90,width=12cm]{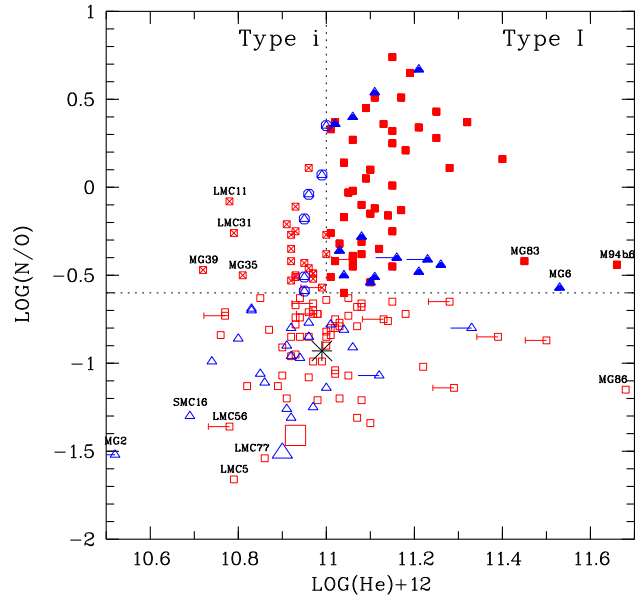}
   \caption{N/O-He/H.
Squares are for \object{LMC} objects and triangles for \object{SMC} ones. 
Filled symbols represent Type$\,$I PNe, and open symbols "standard" PNe. 
The Type$\,$i's (see text) are represented by crossed squares (\object{LMC})
or circled triangles (\object{SMC}). The star marks the solar value. 
The same symbols will be used in all subsequent figures. 
The vertical and horizontal dotted lines indicate 
the separation between Type$\,$I, Type$\,$i, and "standard" PNe.
In all the figures, \object{LMC} objects are plotted in red squares 
and \object{SMC} ones in blue triangles, the color figures being available 
in the electronic form of this paper.}
   \label{nohe}
\end{figure}
\begin{figure}[htbp]
   \includegraphics[angle=-90,width=12cm]{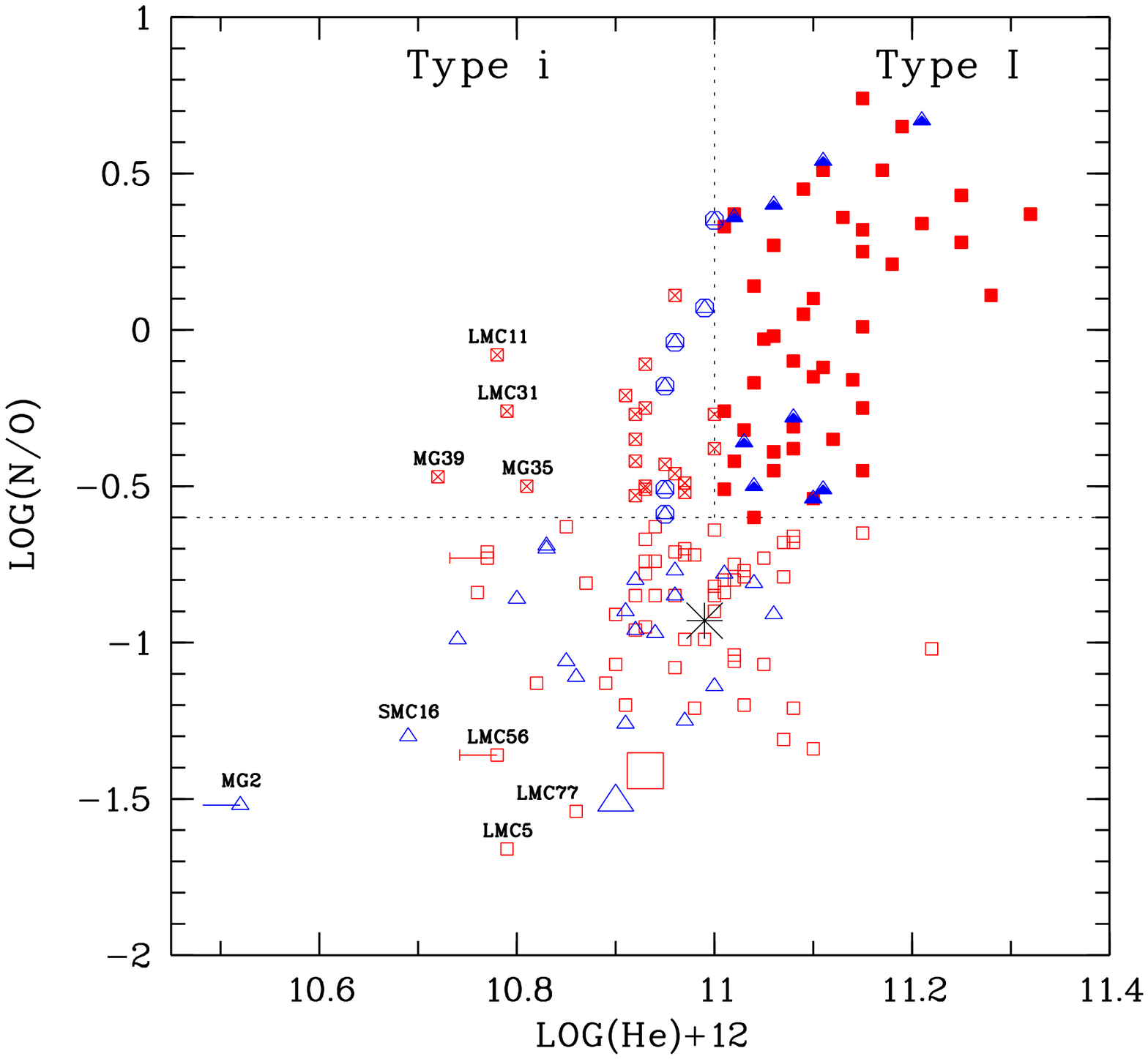}
   \caption{N/O-He/H.  Same as Fig.~\ref{nohe}, but only
with   objects with accurate determinations.  
(see text Sect.~4.1). 
}
   \label{noheNEW}
\end{figure}

\begin{figure}[htbp]
   \includegraphics[angle=-90,width=12cm]{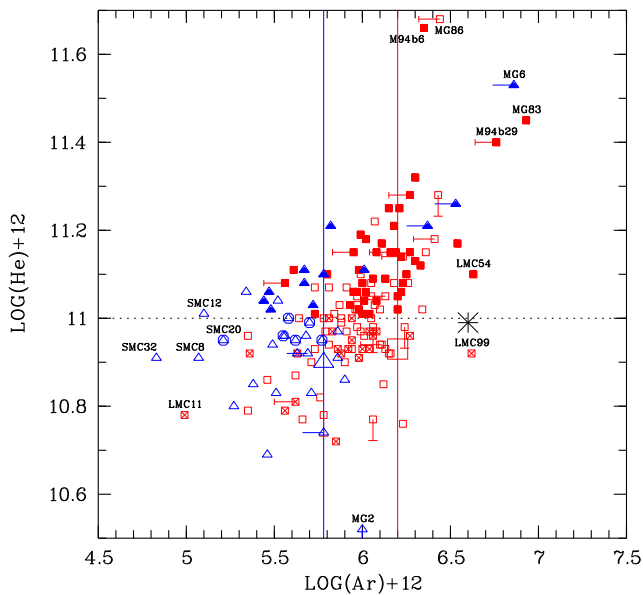}
   \caption{He/H versus Ar/H. The vertical lines are marking the average 
HII regions values in argon for \object{SMC} and \object{LMC}. 
Symbols as in  Fig.~\ref{nohe}. }
   \label{hear}
\end{figure}

\begin{figure}[htbp]
   \includegraphics[angle=-90,width=12cm]{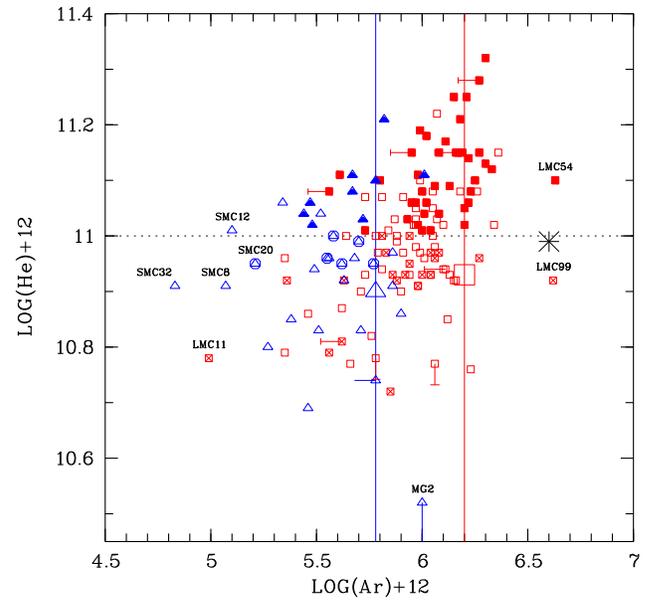}
   \caption{He/H-Ar/H, as Fig.~\ref{hear}, but only with objects with accurate
abundance determinations.}
   \label{hearNEW}
\end{figure}

The N/O ratio increases with  He/H,
however the slope of the correlation is different for the 2 classes.
This correlation is interpreted as due to the mixing of the $2^{nd}$ 
{\it dredge-up} products (He and N) into the envelope.
The Type$\,$i PNe seem to prolong the Type$\,$I PNe correlation, 
(characterized by a larger
helium enrichment, in Fig.~\ref{nohe}) , rather than  the relation for 
the non-Type$\,$I PNe.
On the other hand, it appears that many of them have low initial
abundances, as can be seen in a He versus Ar diagram. 
To show once more that we believe this is a real effect and not due to 
poor abundance determination, we present this 
diagram in the two forms again~: with the full initial sample in 
Fig.~\ref{hear} and after removal of the poorly determined objects in 
Fig.~\ref{hearNEW} (from here onwards, we will then always be only working
on the clean sample of 157 objects). To conclude then for the Type$\,$i PNe, 
the four objects with lowest  He abundance, but N/O above -0.6 
(our definition of Type$\,$i), \object{SMP LMC 11}, \object{SMP LMC 31},
\object{MGPN LMC 39}, and \object{MGPN LMC 35}, 
also have a low argon abundance (Fig.~\ref{hearNEW}, Fig.~\ref{noarNEW}). 
The first three also have a low oxygen and a low neon abundance (Fig.~\ref{neoNEW}).
This may therefore indicate that the major difference between 
Type$\,$I's and Type$\,$i's lies in the lower initial abundances for the latter.

We now discuss the global properties of the N/O versus He/H 
diagram (Fig.~\ref{noheNEW}). Note that two objects  
 with very  low He abundances, 
\object{SMP LMC 26} and \object{LHA 120-N 99} (Monk et al.~\cite{monk}), 
lie outside the diagram, but are very low excitation objects for which a 
good He abundance determination is difficult. 
A few objects have both low He and low N/O~: for instance 
\object{SMP SMC 16}, \object{SMP LMC 5}, \object{SMP LMC 56}, 
\object{SMP LMC 77}. As they also show a low Ar (or O) abundance, these 
are also candidates for progenitors with low initial abundances (and 
therefore older).
However, \object{SMP LMC 77} also has a high carbon abundance 
(unfortunately the only object in this group with a known C abundance)
and can therefore be an object where the processing has gone further than 
the triple-alpha reaction and destruction of He, favored by a low 
initial abundance (as discussed 
in Paper~I). All those objects should therefore also be observed in the UV 
to determine their carbon abundance. 

\begin{figure}[htbp]
   \includegraphics[angle=-90,width=12cm]{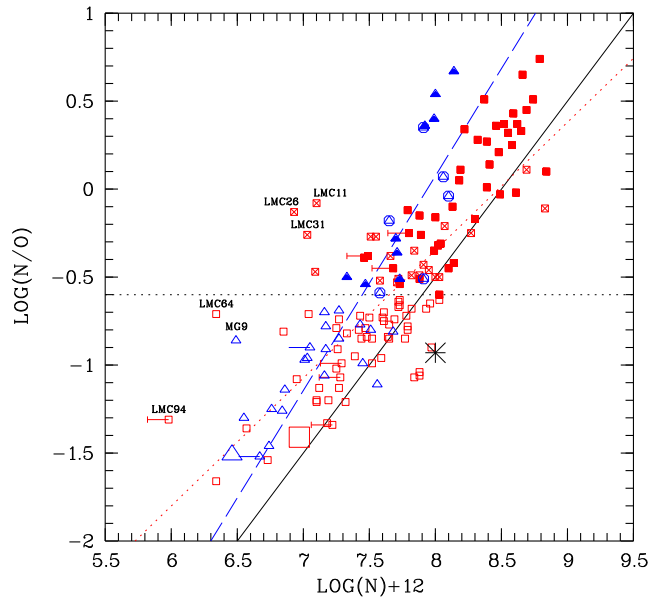}
   \caption{N/O-N/H relation. 
Symbols as in  Fig.~\ref{nohe}.
The dotted line is the fit to \object{LMC} objects, and the dashed line the fit
to \object{SMC} objects. The continuous line is the relation for Galactic objects 
as shown by Jacoby and Ciardullo ~\cite{jacci}. }
   \label{nonNEW}
\end{figure}

A correlation between N/O and N/H is seen in Fig.~\ref{nonNEW}. The large 
sample available here allows us to distinguish  different slopes 
for the two Clouds (N/O=a$\times$(N/H)+b)~:
\begin{itemize}
  \item for LMC, a $\simeq$ 0.727 $\pm$ 0.047  
  \item for SMC, a $\simeq$ 1.220 $\pm$ 0.039  
\end{itemize}
\noindent
These different slopes are an indication that the enrichment processes are
more efficient in metal poor galaxies, as already suggested in Paper~I. \\
The "peculiar" objects  in this diagram are the same ones already discussed
in Fig.~\ref{noheNEW}, with the addition of three more with low N abundance:
\object{MGPN SMC 9}, \object{SMP LMC 64}, \object{SMP LMC 94}.
\object{SMP LMC 64} is a peculiar object with extremely high densities and
temperature, already discussed by Dopita et al.~(\cite{dopSMP64}).  
\object{MGPN SMC 9} and \object{SMP LMC 94} both seem to be low in abundances
for many elements and are therefore  candidates with an old progenitor. 


\subsection{Oxygen}
\begin{figure}[htbp]
   \includegraphics[angle=-90,width=12cm]{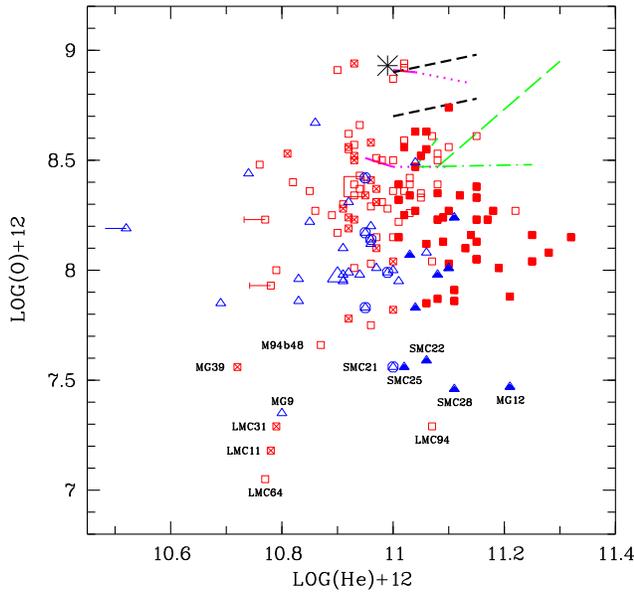}
   \caption{ Relation between oxygen and helium abundances. The various line
 segments are the predictions from the models of 
Marigo et al.~(\cite{mari3}). Upper curves are with galactic metallicities, and
lower with the \object{LMC}.}
   \label{oheNEW}
\end{figure}

In studies of the chemical evolution of galaxies through the analysis of
gaseous nebulae, the oxygen abundance is usually taken as a reference for
the global metallicity and then used as the tracer of the evolution.
When starting this project, it was our intention to follow this practice
and use the oxygen abundance, as measured in PNe, to characterize the
metallicity of the galaxy at the time when the PN progenitor was formed.
While the lifetime of the nebula and its progenitor star is relatively short 
for H\,{\sc ii}~regions compared to the total lifetime of the galaxy, and
hence oxygen there measures the metallicity at the present time (at least
locally) for PNe, this method assumes, however, that no processing of the
initial O abundance has occurred during the life time of the progenitor star.
It rapidly became clear (see Paper~I) that this usual assumption should be
questioned, especially for low metallicity galaxies:
the oxygen abundance measured in PNe does not seem to be very different from the
mean abundance in H\,{\sc ii}~regions, although some progenitors might be 
rather old. 
Looking in more detail at the various diagrams, 
we see that about half of the PNe (and more for the \object{SMC} than
for the \object{LMC}) have oxygen abundances even above this average value,
some of them  with large over-abundances (see for instance Fig.~\ref{oheNEW}).
This fact is already present in earlier data by various authors, but
was never specifically commented on.
The large sample available here allows a detailed discussion of this problem.

The enrichments in PNe are much more apparent in the Magellanic Clouds than
in our Galaxy, which is clear for nitrogen or carbon, but also seems to be 
the case for oxygen.
On average, it appears that the enrichment in oxygen could be as large as  
the amount present at the time the progenitor star was formed.
Therefore the question of oxygen production cannot be neglected anymore.
On the other hand, a fair fraction of PNe have lower oxygen abundances
than the average H\,{\sc ii} value, but it is not clear whether this
reflects the initial abundance directly, or whether it also includes
some destruction during the AGB phases.

Indeed, oxygen abundances can be affected by processing in the PNe
progenitor stellar cores, in at least two ways.
On one hand, oxygen destruction can occur during the CNO cycle in the more
massive progenitors (specifically the ON cycle) and thus affects the 
abundance seen in the nebula. 
On the other hand, nebular abundances, in particular carbon, could be 
substantially modified during the $3^{rd}$ dredge-up by mixing with freshly
core-processed material (Paper~I).
This mixing could also lead to enhancement of the oxygen abundance.
In the next two sections, we look for evidence of such modifications
which, if demonstrated, will require the definition of another metallicity
tracer, unaffected by processing during the life-time of PNe progenitor stars.

\subsubsection{Oxygen destruction}

\begin{figure}
   \includegraphics[angle=-90,width=12cm]{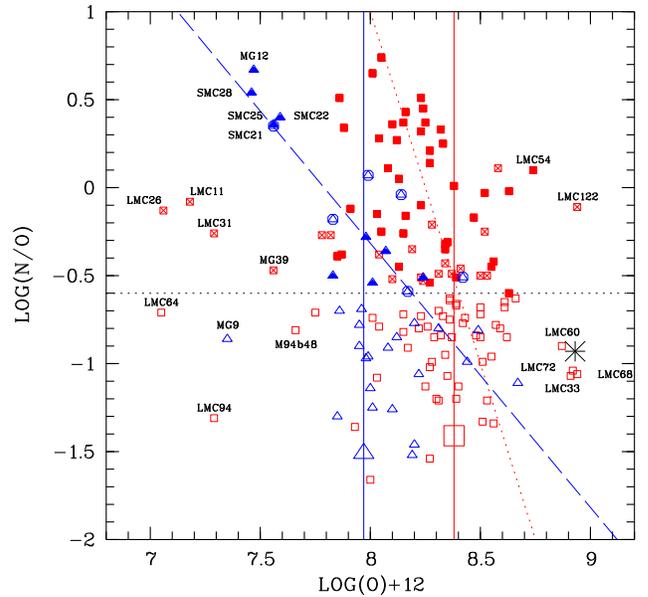}
   \caption{N/O-O/H relation. Symbols as in  Fig.~\ref{nohe}. 
The dotted  (\object{LMC}) and dashed (\object{SMC}) lines are only fits to the 
Type$\,$I nebulae.
The vertical lines underline the HII regions mean abundance values, while the 
horizontal one marks the separation between Type$\,$I  and "standard" nebulae. }
   \label{nooNEW}
\end{figure}
Figure~\ref{nooNEW} presents the N/O versus O relation. 
It appears that about half the \object{SMC} Type$\,$I nebulae display lower
oxygen abundance than average, lower than the  mean value given by the HII~regions. 
This effect is also present for the \object{LMC} objects, but  partly hidden by 
some outliers with higher O abundance, which will be discussed at the end 
of this section (the objects with both low O {\it and} low N/O are the 
same ones as were already singled out in the previous diagrams). 
The overall effect is best seen when introducing the slopes of the relations
for Type$\,$I objects only (Fig.~\ref{nooNEW}). 
The Type$\,$i objects are more dispersed, and will be discussed specifically 
later.

\begin{figure}[htbp]
   \includegraphics[angle=-90,width=12cm]{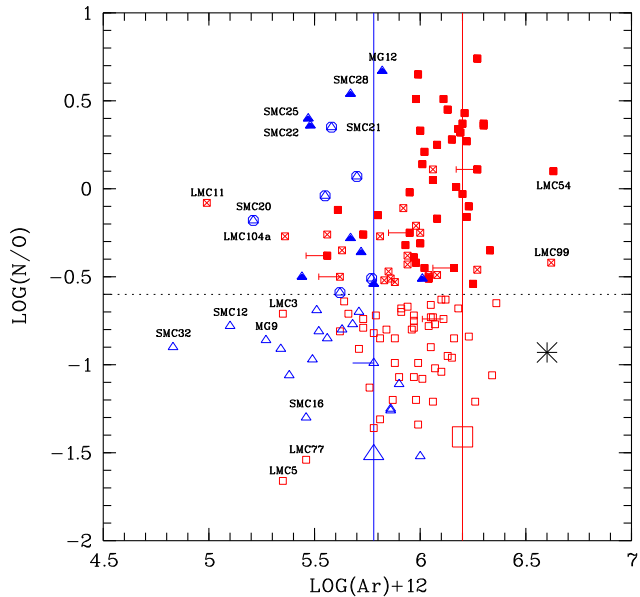}
   \caption{N/O-Ar/H relation. Lines as in Fig.~\ref{nooNEW}.}
   \label{noarNEW}
\end{figure}

As the Type$\,$I nebulae are generally believed to arise from higher mass 
and therefore younger progenitors (Peimbert~\cite{peim3}; see also Stanghellini 
et~al.~\cite{stang1}), this effect cannot be due to a lower initial metallicity. 
This is indeed demonstrated in Fig.~\ref{noarNEW} where Ar is plotted instead of
oxygen in abscissa: the Type$\,$I nebulae are the ones with the largest argon
abundance, which is typical of younger objects.
It is significant that essentially no reliable object has, within the error
bars, a larger argon abundance than in H\,{\sc ii}~regions, contrary to
oxygen (see sect.~\ref{argonsulfur}). 
The two noticeable exceptions are \object{SMP LMC 54} and \object{SMP LMC 99}).
The simplest explanation is that we observe some oxygen destruction, as
predicted by stellar evolution models, especially for the higher masses
where the ON cycle should be more efficient.
 
The observed trend is metallicity dependent, because it is not seen in our Galaxy
and is small in the \object{LMC} and much greater in the \object{SMC}, therefore
implying a metallicity dependence for the corresponding nuclear reactions. 
This dependence, and the large sample of objects used here, indeed helps to
emphasize this effect.
As a consequence of this oxygen depletion, the N/O enhancement is 
therefore even larger than previously thought in Type$\,$I nebulae
(see also Fig.~\ref{oarNEW}, O/H versus Ar/H).

An anti-correlation between N/O and O is also apparent in  Fig.~\ref{nooNEW}, 
and is more pronounced for \object{SMC} objects than for \object{LMC} ones.
This anti-correlation has been known for quite some time (Peimbert~\cite{peim3}) 
and has been discussed more recently, although with a much smaller sample of 
objects, by Costa et al.~(\cite{cost1}). It is also interpreted as due to a 
higher efficiency of the surface enrichment through dredge-ups at lower 
metallicities. Costa et al.~(\cite{cost1}) also argue that the effect is even clearer 
when PNe are grouped by mass range of their progenitor star, but we cannot do 
this here in view of the very few masses really known in the Clouds. A hint 
of this is seen in Fig.~\ref{nooNEW}, as the anti-correlation is 
more pronounced in the upper part of the diagram (Type$\,$I objects) than in the 
lower part populated by "standard", presumably less massive objects. 

In a recent study of ten Galactic objects, Marigo et al.~(\cite{mari3}), in 
discussing an O-He diagram, point out  that 
high He abundance (He/H $> 0.14$ by number, that is log(He) = 11.15) is 
correlated with a low oxygen one. 
While this could be an obvious sign of the onset of the ON cycle 
in the most massive objects, they argue against this hypothesis, because 
their models then predict too high a nitrogen abundance, and prefer to assume 
a lower initial metallicity despite the problems raised by such low 
metallicity in objects close to the Galactic center.
The same diagram for our objects can be seen in Fig.~\ref{oheNEW}, with the 
results of Marigo's models superposed: the situation is more complex and the
models do not reproduce all the observed data well.  
While objects with  both low helium and low oxygen (as seen in the \object{LMC},
in particular) are probably objects with low initial metallicities, the LMC
objects with the highest helium content, log (He) $>$ 11.15, which are all
Type$\,$I's by definition, do indeed present a lower oxygen abundance
than the average non-Type$\,$I.
This cannot all be ascribed to variations in initial composition, or else
it would also be reflected in other elements, such as argon, which is not
the case.
We therefore need to invoke oxygen destruction via the ON cycle.
This is entirely consistent, as the ON cycle is believed to operate only in the
most massive progenitors, i.e. those of Type$\,$I nebulae.
Some objects in Fig.~\ref{nooNEW} have a much higher oxygen abundance than
the HII~regions, such as \object{SMP LMC 54} or \object{SMP LMC 122} (a Type$\,$i).
This implies that we cannot exclude oxygen production, particularly  for 
Type$\,$I and Type$\,$i objects. 
In those nebulae,  the observed oxygen abundance is then the combined result  
of both  destruction and production. 
These two examples probably have intermediate-mass progenitors  
(less than 3~$M_{\sun}$), where  oxygen production should be possible.
\subsubsection{Oxygen enrichment - CO core and thermal pulses}
In Paper\,I we have shown that the carbon abundances are highly enhanced in
the Magellanic Clouds. As discussed there,  
the $3^{rd}$ dredge-up occurs in all PNe, but its efficiency appears to be 
higher with lower initial metallicities.  The N/O versus O anti-correlation
(discussed in the previous section) is interpreted the same way. 
This has also been shown by Costa~et~al.~(\cite{cost1}) and theoretically 
confirmed by Marigo~et~al.~(\cite{mari3}).
This fact offers also an easy explanation for the higher
number of carbon stars found in galaxies with metal-deficient composition.
During the $3^{rd}$ dredge-up, however, the carbon is not the only element
transported. 
In the CO core, oxygen has been  produced by $\alpha$ capture on a 
carbon nucleus $^{12}C (\alpha, \gamma)^{16}O$, although this process has
a smaller probability of occurring than the 3 $\alpha$ process.
It is possible (as suggested by the oxygen abundances observed here, often higher 
than the ones of HII regions) that 
the nuclear reaction rate of this reaction is not as accurately known
as previously thought (Bono~\cite{bono}, or Metcalfe~et~al.~\cite{methan}).

On the other hand, during the thermal pulses, the fusion of Hydrogen produces
$^{13}C$ (and not $^{14}N$) for central stars with core masses smaller than 1~$M_{\sun}$.
Oxygen is then also produced from $^{13}C (\alpha, n) ^{16}O$. 
This reaction is a strong source of neutrons, inducing the {\it s process}:
the observed over-abundances of some high atomic weight elements in AGB stars is
proof of its operation. It would thus also contribute to the production of oxygen.

A  recent model with overshooting (Herwig~\cite{herw1}, Werner \& Herwig~\cite{wern2})
determines layers where the chemical composition is even more enhanced in 
oxygen~ than in the standard CO core composition (by number):
\begin{itemize}
  \item helium : 25-33\% (instead of 22\%)
  \item carbon : 50\% (instead of 76\%)
  \item oxygen : 17-25\% (instead of 2\%).
\end{itemize}
With such a core composition, the $3^{rd}$ dredge-up in operation,
and the strong helium and carbon enrichment observed (up to 100
times for C, or even more), 
one can expect an enrichment in oxygen, too. 
This would be  particularly visible in the Magellanic Clouds, because the 
quantity of metals present in the progenitor star is initially low, and
the enrichment processes are more efficient at lower metallicity.

This therefore seems the easiest explanation for the observed  enhancement 
in oxygen. It may at the same time represent observational evidence  
for the existence of overshooting. When comparing the core composition in 
models with and without overshooting, it also appears that the carbon 
fraction is higher in the latter (the reverse is true for oxygen)~: a 
precise determination of C abundances in more PNe would therefore be a 
key observation confirming this interpretation.  
A simple evaluation can show that a reasonable number of thermal pulses 
(a few tens) are enough to produce the required quantity of oxygen 
to match the observed oxygen abundance and, at the same time, achieve 
the huge enrichment seen in carbon.

Recent semi-analytical models by Marigo et al.~(\cite{mari1}) and 
Marigo~(\cite{mari2}) agree closely with our observations, and with the
possibility of oxygen production (within some initial mass range).
While the oxygen production for an initial solar metallicity is negligible, 
enrichment is predicted at low metallicity (at Z = 0.008, as in the \object{LMC}).
But with an \object{LMC} initial  composition, the oxygen production in
Marigo's models is only strong in the $1.5~M_{\sun} < M < 3~M_{\sun}$ mass 
range. This is consistent with Type$\,$I PNe in the Magellanic Clouds with
initial star masses greater than $3~M_{\sun}$, as no clear oxygen 
enrichment is seen in them. 
It would be desirable to have  such models calculated
for even lower metallicities, such as  \object{SMC} ones, to confirm the trend
observed in our data. 
Charbonnel~(\cite{charb}) emphasizes  that the surface abundance of 
$\element[][16]{O}$ in massive progenitors is the result of the 
competition between efficiencies of the third dredge-up (for production) 
and hot-bottom burning (for destruction), but that rotation also 
significantly enhances the surface abundance following the second dredge-up, 
especially at low metallicities. 
We can  provisionally  conclude that, as shown by the models, 
the explanation of the large observed enrichments  lies both in the
lower initial metallicities and in the corresponding increase in duration 
and efficiency of the phases of thermal pulses at the end of the AGB stage.

We note that the objects where oxygen production has probably been strong,
such as those with much higher oxygen abundance than in the HII regions
(as seen, for instance, in  Fig.~\ref{oarNEW}~: \object{SMP LMC 122}, 
\object{SMP LMC 33}, \object{SMP LMC 60}, \object{SMP LMC 68}, or 
\object{SMP LMC 72}) are generally not Type$\,$I objects. 
They stand out in Fig.~\ref{nooNEW} because of their low N/O (presumably 
due simply to their high O), but they do not show signs of peculiar N or 
He production, and should therefore have progenitors with lower masses
where the HBB was not active.
A carbon-abundance determination would be helpful for reaching a conclusion. 

Globally, production of oxygen (for all masses) and destruction of it (for the 
higher masses only) compete for the net result;
the enrichment appears to be maximal in an intermediate mass range 
(typically 1.5 to 3~$M_{\sun}$ for an \object{LMC} composition), but the net
effect for higher masses is destruction of oxygen.
When looking then at the Type$\,$i objects, which are rather dispersed in 
oxygen abundance in Fig.~\ref{nooNEW}, the consequence from the above is that the 
ones with higher oxygen abundances (\object{SMP LMC 54} 
and \object{SMP LMC 122}, right side of the diagram) should have lower 
progenitor masses than the ones on the left side, such as \object{SMP LMC 11} or  
\object{SMP LMC 26}, where oxygen destruction has been dominant.

\subsection{Argon and sulfur}
\label{argonsulfur}
The only other elements whose abundances can be determined easily from 
optical spectroscopy are argon and sulfur. 
In spite of the weakness of the observed lines, argon is a good element
for use in the determination of the initial composition of PNe central 
stars, so it should not be affected by transformations during the AGB phase. 
\begin{figure}[htbp]
   \includegraphics[angle=-90,width=12cm]{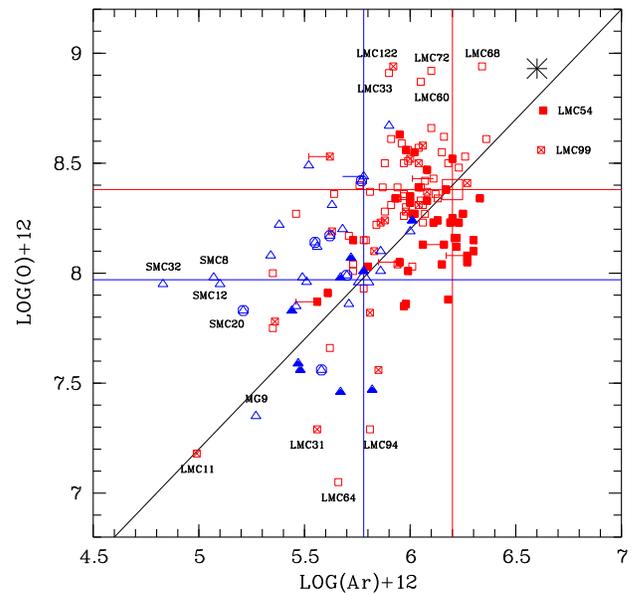}
   \caption{O/H-Ar/H relation.  
The diagonal is the line with constant O/Ar going through the \object{SMC}
and \object{LMC} HII regions values, which are underlined by the vertical and 
horizontal lines. }
   \label{oarNEW}
\end{figure}

Figure~\ref{oarNEW} shows that indeed only a few objects exhibit a larger argon
abundance than in H\,{\sc ii}~regions  (contrary to  oxygen). The points 
are grouped well around the line of slope 1 fitted to the 
H\,{\sc ii}~regions and solar values. The average uncertainty on argon 
abundances should not exceed 0.2 dex by much, and the few points with 
values significantly above the  H\,{\sc ii}~regions all correspond to peculiar
objects (\object{SMP LMC 54} and \object{SMP LMC 99}), as already pointed
out in previous diagrams.
Furthermore, the Type$\,$I PNe all have abundances close to the 
H\,{\sc ii} ~regions'mean value, confirming that these objects are
massive and young. 
Argon could therefore be used instead of oxygen as a good tracer of
chemical evolution or time. 

\begin{figure}[tbp]
   \includegraphics[angle=-90,width=12cm]{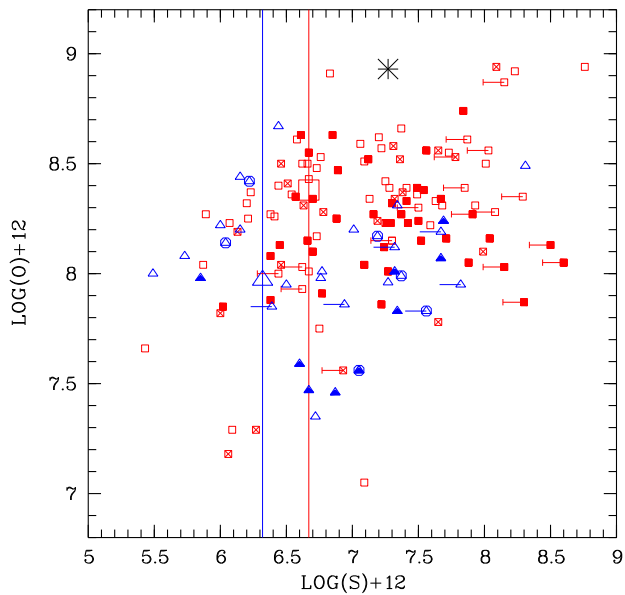}
   \caption{O/H-S/H diagram with all the objects. Symbols and lines as in 
previous figures. }
   \label{osNEW}
\end{figure}

\begin{figure}[htbp]
   \includegraphics[angle=-90,width=12cm]{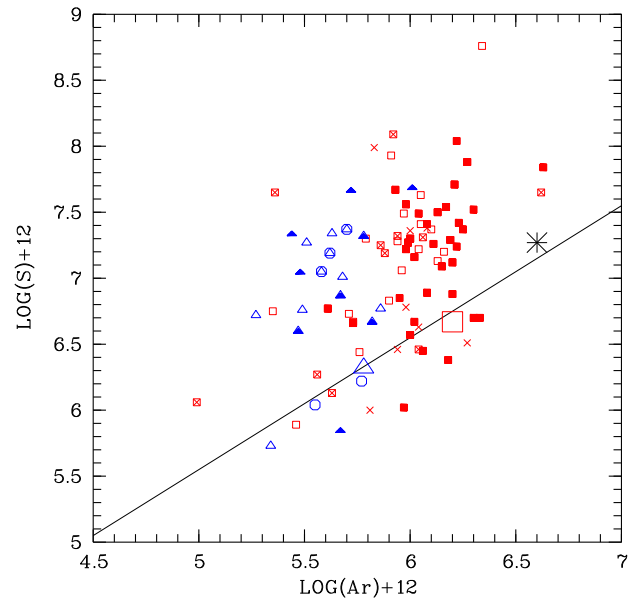}
   \caption{S/H-Ar/H diagram with only the objects with best S 
determination. Objects with upper limits  in  both elements are not plotted. 
The continuous line is a slope 1 line going through the HII regions, for 
reference.  }
   \label{sarNEWgood}
\end{figure}

The same should be true for sulfur, at least from a nucleosynthesis point of view.   
In Fig.~\ref{osNEW}, a broad correlation appears between oxygen and sulfur,
but with higher dispersion than for argon.
Here many objects seem to have high abundances when compared to H\,{\sc ii}~regions. 
\begin{figure}[tbp]
   \includegraphics[angle=-90,width=12cm]{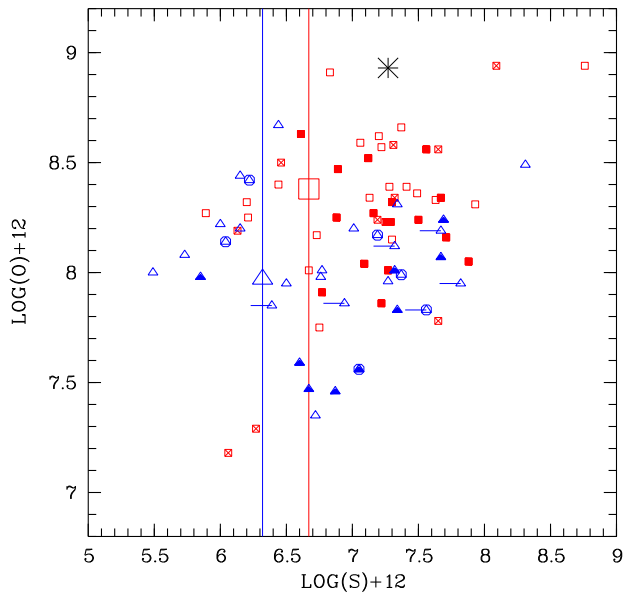}
   \caption{O/H-S/H diagram as Fig.~\ref{osNEW} with only the objects
with best S determination (see text).}
   \label{osNEWgood}
\end{figure}
This can, however, not be taken at face value, as the  accuracy of sulfur
abundance determinations is affected by well known problems.  
In particular, for high excitation objects like many PNe, the contribution 
of [SIV] lines to the abundance determination is essential 
(see, e.g., Dennefeld and Stasinska~\cite{denn2}) but is not available here. 
For many objects, we do not even have a detected [SIII] line. 
Adding the fact that the only available [SIII] line (in our observed range), 
the $\lambda_{6312}$\,\AA ~line, is often too weak or blended with
an oxygen ($[\,OI$]) line and that an electronic temperature in the sulfur
zone should be derived as well for the ICF method, we end up with a rather
large inaccuracy in the sulfur abundance determination.
An effort is underway to improve this by complementary observations 
(Leisy~\&~Dennefeld, in preparation). 
For the moment, we can try to improve the picture by taking only objects 
where both [SII] and [SIII] are available (as in Fig.~\ref{osNEWgood}).
Then about only half of the PNe are usable,  the  dispersion in the sulfur
diagrams is reduced, and most objects with large ICF also disappear. 
But the sulfur abundance determination  is still not good enough 
to use sulfur as the metallicity indicator.  
One would for instance expect a good correlation between S and Ar, but  
Fig.~\ref{sarNEWgood}, where a line of unit slope is plotted for reference,
shows that the correlation is still not as good as expected, 
undoubtedly due to the poor sulfur abundance determination.
Sulfur can therefore not yet be taken here as an element to replace oxygen
as a tracer of the chemical evolution with time, although it is produced 
basically by the same progenitor stars as oxygen.   This can be improved 
in the future by complementary observations in an extended spectral range. 
\subsection{The oxygen-neon relation}
\begin{figure}[htbp]
   \includegraphics[angle=-90,width=12cm]{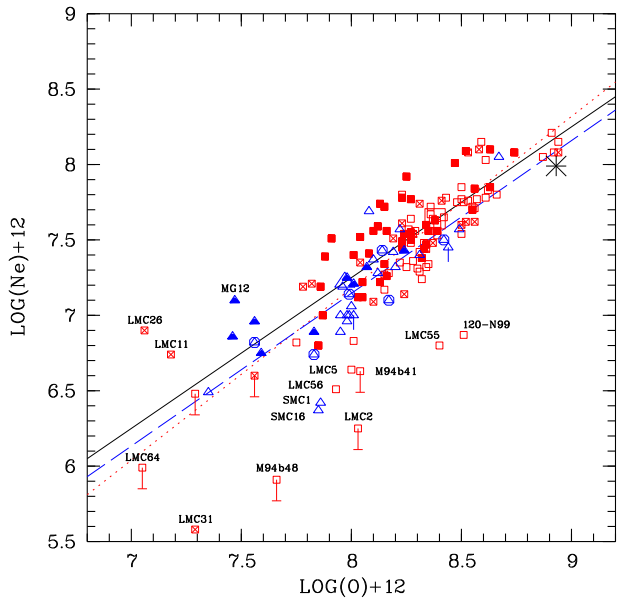}
   \caption{ Relation between neon and oxygen abundances. 
The dotted (slope 1.13) and dashed (slope 1.01) lines are the fit to
the \object{LMC} and \object{SMC} objects, respectively, while the
continuous one is a slope 1 line going through the HII~region values. }
   \label{neoNEW}
\end{figure}

Figure~\ref{neoNEW} shows the Ne/H versus O/H diagram, where a good 
correlation can be seen between these two elements. 
The full line is a unit slope line going through the  H\,{\sc ii}~region 
points only, along which all the PNe points are well-grouped.
A relation between O and Ne was already mentioned, e.g. in our Galaxy by
Henry~(\cite{hen1}), who interpreted it as a sign that these two 
elements were not significantly altered by nucleosynthesis in PNe.  
The relation is much better defined here in the Clouds alone, because of the
larger number of objects available, but is however more difficult to understand
now in view of the oxygen variations discussed in the previous sections. 

\begin{figure}[htbp]
   \includegraphics[angle=-90,width=9cm]{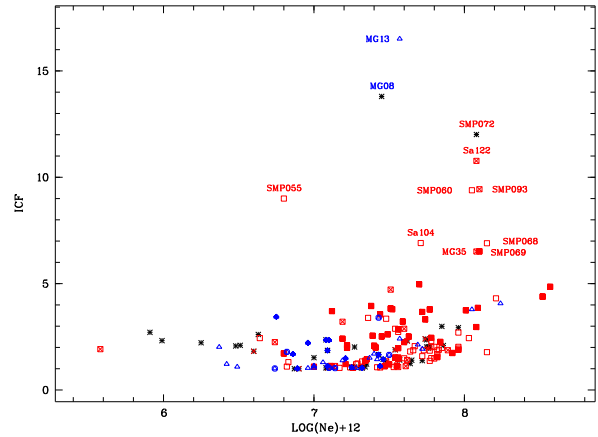}
   \caption{Ionization correction factor (ICF) for neon. Symbols as in 
previous figures; stars indicate objects with poor determinations. }
   \label{icfne}
\end{figure}

The main correlation seen in  Fig.~\ref{neoNEW} is not due to an ICF bias,
because there is no obvious relationship between the (oxygen-dependent) ICF
and the neon abundances, as can be seen from  Fig.~\ref{icfne}.
Only one object with a high ICF,  \object{SMP LMC 55}, falls off the main
relation in Fig.~\ref{neoNEW}, but its neon abundance is too low, not high.

The Type$\,$I objects are slightly offset from the mean relation in 
Fig.~\ref{neoNEW}, as already noted by  Henry (~\cite{hen1}), 
with a more pronounced difference at the low metallicity end.
This offset can be understood as due to some oxygen destruction in the most
massive objects with larger efficiency at lower metallicities, as discussed
earlier.
A clear example is the Type$\,$I object \object{MGPN SMC 12}.
%
The correlation seen in Fig.~\ref{neoNEW}   
implies that, if the oxygen abundance is enriched  by some process, 
the neon abundance has to be enriched, too, to follow the correlation with oxygen.
This is particularly true for those PNe with higher abundances than in
H\,{\sc ii}~regions. It can be understood with  standard nuclear reactions 
%
such as $\element[][14]{N} (\alpha, \gamma) \element[][18]{O} (\alpha, \gamma) 
\element[][22]{Ne}$ where 
the newly produced $^{14}N$ is almost completely converted into
\element[][22]{Ne} in a rich $\alpha$ environment.
Neon could eventually be destroyed to produce Magnesium via the reaction 
$\element[][22]{Ne} (\alpha, n) \element[][25]{Mg}$, but this reaction has a
very low efficiency (1\%) unless the final core mass is greater than 1~$M_{\sun}$. 
As the majority of PNe central stars have core masses below this limit, the
 freshly produced  neon should therefore not be destroyed, except in the 
most massive progenitors.
\begin{figure}[htbp]
   \includegraphics[angle=-90,width=12cm]{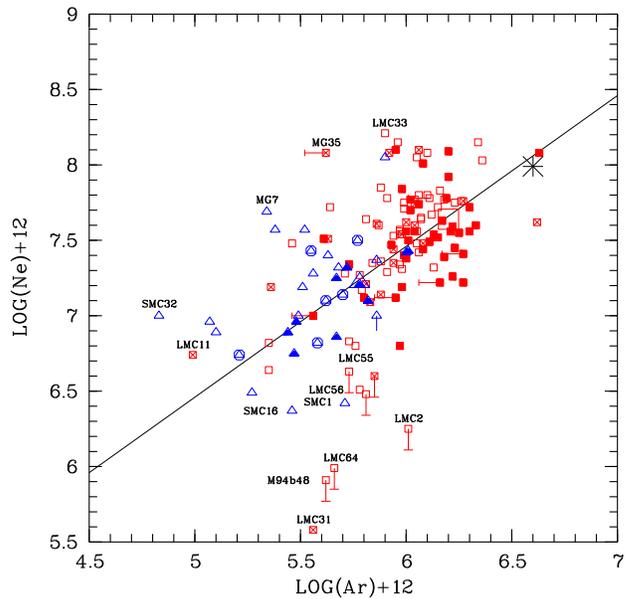}
   \caption{Ne/H-Ar/H relation. The continuous line goes through the HII 
region values. }
   \label{nearNEW}
\end{figure}


Alternately, the neon production could directly correlate with oxygen 
if the $\element[][16]{O} (\alpha, \gamma) \element[][20]{Ne}$ reaction 
were at work, but the reaction rates should then be higher than is actually
believed. 
The fact that neon production is linked to oxygen is strengthened by the
absence of correlation with other elements;
e.g. in Fig.~\ref{nearNEW}, the Ne-Ar relation displays a 
large dispersion where Ne varies over two and a half dex, while Ar varies 
over slightly more than one dex only.
There is indeed no theoretical link between neon and argon productions,
and argon is not expected to be produced or destroyed in PNe either.
It also seems that the O-Ne relation is tighter for higher abundances, which 
is clearly seen for \object{LMC} or \object{SMC} objects in Fig.~\ref{neoNEW}. 
This relation is followed well by all the M31 PNe (with the exception of
precisely one deficient, halo object;  Jacoby \& Ciardullo~\cite{jacci}), 
and a similar picture is found for Galactic objects, except for the halo
ones again (Henry~\cite{hen1}). 


We have aimed at a simple evaluation of the enrichment for neon, 
as done previously for oxygen.
The result is similar: i.e. after a hundred thermal pulses, 
the surface abundance of neon is substantially increased, and  the
oxygen-neon correlation is maintained, even with oxygen (and thus neon) 
production. 

The production of neon in the core does not necessarily follow that of oxygen
systematically, it is very dependent on the 
initial mass of the progenitor. The most plausible explanation for the objects 
lying below the main correlation and in the lower left corner of  
Fig.~\ref{neoNEW} is that they have experienced some  oxygen production not 
followed by neon production. Their progenitors are probably of low mass, with 
low initial metallicities, as also shown by their low argon 
abundance.
Good examples are \object{SMP LMC 5} or \object{SMP LMC 55}, which fall on 
the main correlation in  Fig.~\ref{nearNEW}, with low abundances, and  
show an excess in oxygen. Other objects, such as  \object{SMP SMC 1}, 
~\object{SMP SMC 16},  ~\object{SMP LMC 2}, or ~\object{SMP LMC 56}, show 
a deficiency in neon without peculiarities in either oxygen or argon, which can
be due either to some neon destruction or to a particular initial composition and 
which needs to be studied in more detail. 

Model calculations are clearly desirable for checking the various 
possibilities in detail.
Production of neon in a given range of initial masses is, for instance,
foreseen in the models of Marigo et al.~(\cite{mari1}) and Marigo~(\cite{mari2}),
but the details of the relation, the amount produced, and the precise  
dependence on initial masses, were not explained.
More recently, Marigo et al.~(\cite{mari3}) had to invoke a significant neon 
production (with a correlative low efficiency of its destruction mechanism, 
i.e. a significant reduction in the production of Magnesium and neutrons via 
the $\element[][22]{Ne} (\alpha, n) \element[][25]{Mg}$ 
reaction) to reproduce observed abundances in a small sample of Galactic PNe 
with presumably low, ~\object{LMC} type, initial abundances.  
Effective production of neon in the He-burning shell is also demonstrated by
observations of PG1159 stars (post-AGB stars after a  helium-shell flash; 
Werner et al.~\cite{wern1}). 
\begin{figure}[htbp]
   \includegraphics[angle=-90,width=12cm]{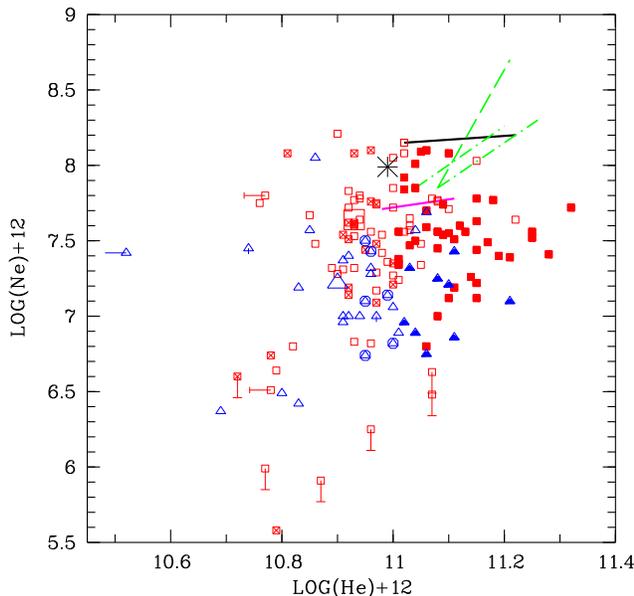}
   \caption{ Relation between neon and helium abundances. The various lines 
represent model predictions from Marigo et al.~(\cite{mari3}); in particular 
the two continuous, almost horizontal lines are for Galactic and \object{LMC} 
abundances.  }
   \label{neheNEW}
\end{figure}
The observational evidence is therefore growing, and theoretical work follows:
increased production efficiency at very low metallicity is shown by
Herwig~(\cite{herw2}).
Specific models to reproduce abundances in PG1159 type objects (bare 
planetary nebulae central stars, Werner \& Herwig, ~\cite{wern2}) show that significant amounts of $\element[][18]{O}$ 
and $\element[][22]{Ne}$ 
from the He-burning shell are brought to the surface by the third dredge-up, 
when the initial stellar mass is below about 1.5~$M_{\sun}$.
For higher masses, some of the neon is burned again into magnesium.  
Recent models by Charbonnel~(\cite{charb}) confirm the production of 
 $\element[][22]{Ne}$ in PNe progenitor stars and show that its abundance 
is a clue to the efficiency of various processes, including the production 
of s-process elements, which are unfortunately not easy to measure in PNe. 
 Clearly more models are required,  
with the specific initial metallicities relevant to the Large, 
and more importantly the Small, Magellanic Cloud. 
As is evident, for instance, in Fig.~\ref{neheNEW} where Marigo's model 
predictions have been superposed to the observed data,
the models of Marigo et al.~(\cite{mari3}) with their set of 
abundances (mostly \object{LMC} abundances, or higher) provide too high values   
compared to the observed range of abundances.

\subsection{ Initial composition}
Following the above discussions, oxygen cannot be used anymore as a tracer 
of the initial composition of a PNe progenitor star (nor should neon). 
Therefore, following a cautionary principle, the  only usable elements are  
those that are believed not to be altered during the various AGB 
phases, and one should then take their sum to define an "initial 
metallicity". Among the elements available in our spectra,  this therefore 
leaves only argon and sulfur (but others such as chlorine could be added later). 
As the accuracy of sulfur abundance determinations is, however, far from 
satisfactory at the moment, we have no other choice but to use argon 
alone as the tracer at this stage (even if we are sometimes limited by the 
faintness of some argon lines). The previous diagrams with argon can thus be 
used to discuss the evolution and to try to identify objects with peculiar
abundances.

We again stress (Fig.~\ref{hearNEW})
the difficulty to make a clear separation between Type$\,$I and non-type I objects. 
What is even more interesting is the 
spread in "total" abundance, showing indeed that the various progenitors 
did not all start with  the same initial abundances. This of course makes the 
definition of an "enrichment" (and hence the definition of a Type$\,$I object) 
the  more difficult.

The N/O versus "total" abundance diagram (Fig.~\ref{noarNEW}) helps 
to clarify this.
In this diagram, the nature of what we have called "Type$\,$i" appears more 
clearly.
While some of them have a "total abundance" that is only marginally 
different from the classical Type$\,$I objects (and were therefore omitted from 
this class only because they are lying close to, but on the "wrong" side of, the 
border line), the majority of them are, indeed, simply 
objects with very low initial abundances.
Then, although the enrichment in He might be high, it is not enough to
make them fall above the He/H limit (which was defined from Galactic
objects, and corrected here for the difference in metallicity between the 
Galaxy and the Clouds, see Paper~I) 
to be called "true" Type$\,$I objects.
This reinforces our claim that the Type$\,$I PNe in the Clouds can
only be defined well once the initial abundance is known, and hence
the true relative enrichment can be calculated.

The other diagrams help to select objects of particular interest, either 
because they have low metallicities, and are then tracers of past chemical 
composition of the Clouds or because they display signs of oxygen processing 
(production or destruction) and are key targets for helping refine the models of 
stellar evolution.
If some objects exhibit both low argon {\bf and} low oxygen (Fig.~\ref{oarNEW}), 
they are candidates for old, hence low-mass, progenitors. This is the case for 
\object{SMP LMC 11} (which is also He-poor) and \object{SMP LMC 31}, both of 
Type$\,$i, and, among the non-Type~I PNe, we mention \object{MGPN SMC 9},   
\object{SMP LMC 26} and \object{SMP LMC 94}.
Finally, SMP LMC 64 is a very peculiar object, with extremely high
temperature and density (Dopita et al.\cite{dopSMP64}), hence with more
uncertain abundance determinations.

If objects appear to be weak in oxygen, but 
{\bf not} in argon, it is a sign of oxygen destruction. 
Notable are \object{MGPN SMC 12}, \object{SMP SMC 21}, \object{SMP SMC 22}, 
\object{SMP SMC 23}, and \object{SMP SMC 25}, all of 
them Type$\,$I or Type$\,$i. This is consistent with the idea that Type$\,$I's 
have high-mass progenitors, where the oxygen destruction is likely to occur.

Objects appearing weak in argon, but not in oxygen are likely to have 
produced some oxygen. This is the case for \object{SMP SMC 8}, 
\object{SMP SMC 12}, \object{SMP SMC 20}, or \object{SMP SMC 32} in 
the \object{SMC};
or in the \object{LMC}, for {SMP LMC 122}, {SMP LMC 33}, {SMP LMC 60},
{SMP LMC 68} or {SMP LMC 72}.
Such objects cannot directly appear as Type$\,$I's  because 
their relatively high O abundance prevents their N/O to cross the Type$\,$I
limit (even with a limit revised to -0.6 for the Clouds). Yet some of them 
nevertheless have a (relatively) high  He abundance. 
That objects in the last two categories are more easily identified in
the \object{SMC} than in the \object{LMC} is the 
consequence of the higher efficiency of the various processes  involved 
(as discussed earlier, and in Paper~I) when the initial metallicity is low.
Finally, more objects with oxygen processing  can be selected from the Ne-O diagram 
(Fig.~\ref{neoNEW}), for instance
\object{MGPN SMC 12}, {SMP LMC 11}, or {SMP LMC 26} (oxygen destruction), 
 or {SMP SMC 1}, {SMP SMC 16}, {SMP LMC 5}, 
\object{SMP LMC 31} or \object{SMP LMC 55} or {SMP LMC 56}
 (oxygen production). \object{LMC 120-N 99} enters in the same category, but 
being a very low excitation object, the determination of its  physical 
parameters is more uncertain. 
Those objects will be subject to  
further observations to refine their diagnostics and derive the properties of 
their progenitor stars (Leisy~\&~Dennefeld, in preparation).

\section{Spatial distribution and chemical evolution}
\begin{figure}[htbp]
   \includegraphics[angle=-90,width=9.5cm]{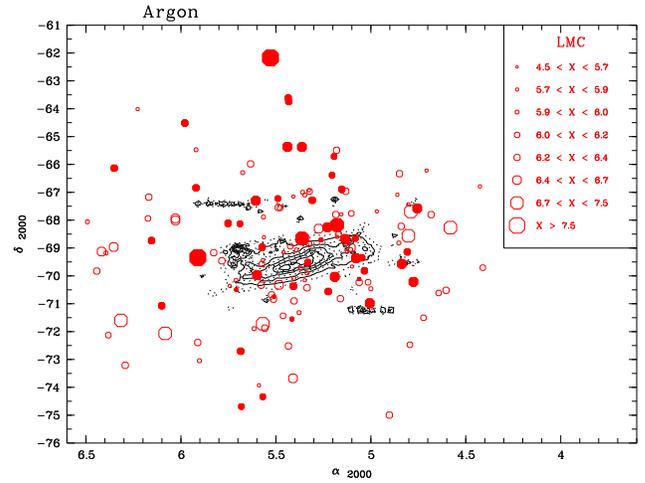}
   \caption{The spatial distribution of PNe argon abundances in the \object{LMC}. 
Each PN is represented by a circle, size whose is proportional
to the total abundance. Filled symbols represent Type$\,$I PNe. }
   \label{lmc-coorAR}
\end{figure}

\begin{figure}[htbp]
   \includegraphics[angle=-90,width=9.5cm]{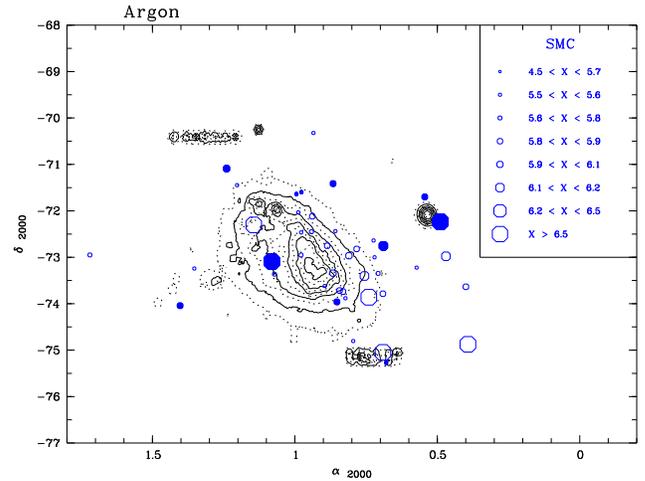}
   \caption{Same as Fig.~\ref{lmc-coorAR}, but for the \object{SMC}. }
   \label{smc-coorAR}
\end{figure}

This very large sample of objects (156 good spectra out of 183 PNe) 
should allow us to check the chemical homogeneity of the Clouds much better 
than with  H\,{\sc ii}~regions, where only about 20 abundance determinations 
are available.  We plot the spatial distribution of our "total" abundances 
(in this case, argon only) in Figs.~\ref{lmc-coorAR}~and~\ref{smc-coorAR}
for the \object{LMC} and the \object{SMC}, respectively. 
We see that contrary to the common assumption derived from the  
H\,{\sc ii}~regions, the abundances seen in the Magellanic Clouds are
not homogeneously distributed, and the Type$\,$I's do not correlate 
with known regions of recent SF.
Neither in the \object{LMC} nor in the \object{SMC}.


Even if we only consider the Type$\,$I PNe, whose abundances should, owing to
their young age, reflect a value close to present abundances, comparable to
H\,{\sc ii}~regions, the Clouds do not appear well-mixed. 
In the \object{SMC}, the abundances seem to divide into two zones: one in the 
South-West with higher values, and one in the North-East with lower ones. 
This would be consistent with a younger age for the south western part of 
the \object{SMC} bar, but somewhat in contradiction with the recent idea that 
SF has proceeded rather continuously in the \object{SMC} over the
past 10 billion years or so (Van den Bergh~\cite{vdb}), a period of time 
longer than the age of the oldest PN progenitor.

For the \object{LMC}, the situation is more complicated. 
Low-abundance objects  seem to be distributed well over the surface of the galaxy, 
such as a halo-type population. However,  
PNe with low mass progenitors would, in their vast majority, trace
SF in an age interval of only about 0.5 to 10 billion years ago,
which corresponds to a period where the stellar formation in
the \object{LMC}, as determined by cluster metallicities, 
was believed to be close 
to zero (Da Costa~\cite{daco}{\bf )}. 

However, this contradiction is only
apparent as studies of field stars (instead of clusters) show that SF was going on 
rather continuously in the past, with a marked increase around 4 billions
years ago  (Van den Bergh~\cite{vdb}), so that the PNe could help determine
the intensity of this SF "renewal".
One would expect more objects with very low metallicities, well-distributed 
over the "halo", but this is certainly an observational bias, as those PNe 
are old and faint and have not been observed in large numbers with 4m class 
telescopes. They are clear targets for larger telescopes.
\begin{figure}[htbp]
   \includegraphics[angle=-90,width=9.5cm]{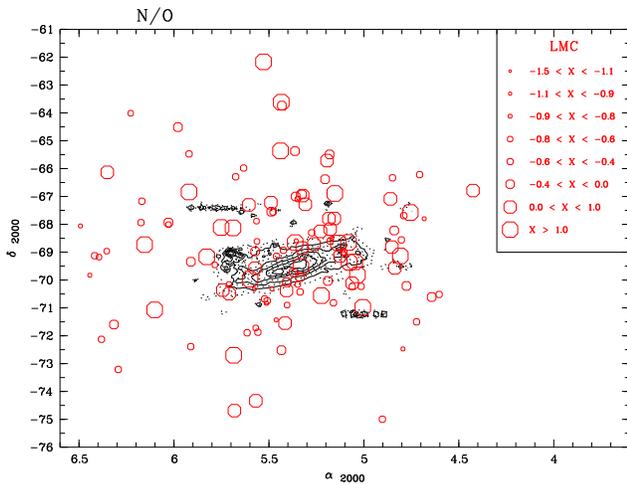}
   \caption{ Spatial distribution of N/O in \object{LMC}, as measured in  PNe. 
The size of the symbol increases with N/O value (details in the insert). }
   \label{lmc-coorNO}
\end{figure}
\begin{figure}[htbp]
   \includegraphics[angle=-90,width=9.5cm]{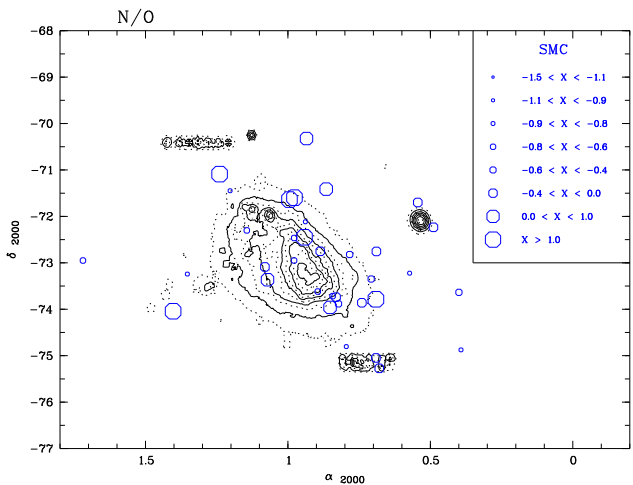}
   \caption{Same as Fig.~\ref{lmc-coorNO}, but for the \object{SMC}.}
   \label{smc-coorNO}
\end{figure}

The higher abundances in PNe seem to be more confined 
to the bar and south eastern region, not correlated with sites of recent 
SF as seen from the H\,{\sc ii}~regions. However, as 
H\,{\sc ii}~regions only trace the very massive stars, this could 
be consistent with the idea that SF is proceeding by 
bursts and rapidly migrating from one place to another. 
The PNe with high abundances would then trace regions of "relatively recent" SF, 
that is, places where the SF has been active recently, but is not ongoing.
Indeed, when looking at the enrichment produced 
by PNe, with the spatial distribution of N/O (Figs.~\ref{lmc-coorNO} 
and ~\ref{smc-coorNO}), it is apparent that 
Type$\,$I PNe are distributed uniformly, so that "recent" SF was proceeding
throughout the Clouds.
The necessary test for the chemical homogeneity, both in space and time, is 
to compare the abundances derived from a given PN with those derived from 
an H\,{\sc ii}~region located in the same physical region. More 
H\,{\sc ii}~regions therefore need to be observed.
\section{Conclusions}
We have presented new spectra of 65 Planetary Nebulae in the Magellanic 
Clouds, for which abundances were derived. We reanalyzed 
published data of other PNe in the same way to produce a homogeneous 
sample of abundances for 183 PNe in both Clouds, out of which 
156 have been finally retained for further analysis, being considered 
as having  good quality data. 
This large sample allows us to highlight several observational facts for the
first time through analysis of classical abundance-abundance diagrams. 
We have arrived at the following conclusions: 
\begin{itemize}
  \item There is a continuity in He and N/O between Type$\,$I's and non-type 
I PNe, so that the distinction between the two types becomes difficult. 
   \item The  third dredge-up is taking  place in most objects (as discussed 
  in  Paper~I)
  \item  The Hot Bottom Burning takes place in the most massive objects, 
  leading to high N/O values.
  \item Overshooting seems to be active and is a good way to explain 
the observed enhanced oxygen abundance: it leads to a large oxygen
production in the core, later released during the thermal pulses. 
Inclusion of rotation in models seems, however, to strongly affect the
oxygen production, too, especially at low metallicities. 
  \item Many reactions are more efficient at lower initial metallicities.  
This is true for the HBB, which then takes place in lower mass stars (as seen 
in the \object{SMC}). It is true for the 3rd dredge-up, taking place in nearly
all stars in the \object{SMC}, 50\% of them in the \object{LMC} and only a
small fraction in our Galaxy. It is true for 
     oxygen destruction or production,  not seen in our Galaxy, but seen
in the \object{LMC} and even more in the \object{SMC}. 
  \item  Abundances of oxygen and  neon in PNe envelopes appear to be  
    affected by internal processing and do not reflect, contrary to what was 
believed earlier,  the initial 
abundances of the progenitor star. 
\end{itemize}

As a consequence, neither oxygen nor  neon should be used to trace the 
chemical history of a galaxy when using PNe. 
Chlorine, sulfur, or argon, which should not be affected by processing 
in the PNe progenitor star, should instead be used to measure the  
initial abundances of the progenitor. 
Unfortunately the first element, chlorine, has very weak emission lines, and in
no object in  \object{LMC} or  \object{SMC} is the $[Cl\,III]$ doublet even detected.
The abundance of the second, sulfur, is presently not well-determined, 
because some of the lines of the dominant ions are too weak, or they fall outside
the classical optical range thereby making the uncertainty in the final
abundance very large.

As a result only argon is presently suitable and it is used as the 
indicator for  the "initial metallicity" of  PNe  progenitor stars, instead 
of an average of several different elements. 
Observations of some of the brighter PNe with larger telescopes to measure 
fainter lines and an extension into the IR (to measure additional 
ionization stages of elements) should help to overcome this limitation. 
Similarly, access to the UV range is needed to obtain abundances of elements 
such as C, which play a key-role in understanding on-going processes. 
It should also be noted that a fair fraction 
of already  published spectra are of poor quality, 
often not allowing even for a proper temperature or density determination.
This increases the scatter in the diagrams and reduces the ability to 
interpret some of the observed effects. An observational effort to secure more high-
quality spectra is therefore appropriate. 
Some effort for fainter objects (fluxes less than 100 
$10^{-16}\;\mathrm{erg}\;\mathrm{cm}^{-2}\;\mathrm{s}^{-1}$)
would at the same 
time reduce the existing bias in favor of brighter (higher mass?) objects.
This, together with observations of some isotopic ratios whenever 
possible (such as $^{12}C$/$^{13}C$), 
would then allow finer comparison between observed abundances and 
stellar evolution predictions and fully exploit the potential of PNe as 
tracers of the chemical evolution in Local Group galaxies. 

\begin{acknowledgements}
PL would like to thank his parents for hospitality and support during 
a critical period between the start of this work several years 
ago and its final completion. MD acknowledges  ESO's support, through its 
short-term visitor program in Chile, where part of this work was completed. 
We thank the referee for useful comments that helped to improve the
clarity of this paper.
\end{acknowledgements}
 
\appendix

 \section{Fluxes}
We present here the  tables of intensities for all  65 PNe observed by us.
For completion, all objects are listed, including those already discussed in Paper~I.
In addition, for practical reasons, to show what  intensities were used in 
our new, homogeneous, abundance determinations, we present intensities also for  
the full sample (183 objects), all in the same format.

These tables will  be accessible only in electronic form, at the CDS in Strasbourg.
\clearpage
\newpage


\begin{thebibliography}{}
\bibitem[2001]{alle1} Allende Prieto, C.; Lambert, D.L. \& Asplund, M. 2001, 
	ApJ 556, 63 
\bibitem[(1979)]{bec1} Becker, S.A. \& Iben, I., Jr. 1979,
	 ApJ 232, 831
\bibitem[(1980)]{bec2} Becker, S.A.\& Iben, I., Jr. 1980,
	 ApJ 237, 111
\bibitem[2000]{bono} Bono, 2000, Private communication
\bibitem[2005]{charb} Charbonnel, C. 2005,  PNe as astronomical tools, Gdansk 
	conference, R. Szczerba, G. Stasi\'nska \& S. K. G\'orny Edts, 
	AIP Conf. series, in press
\bibitem[2000]{cost1} Costa, R.D.D.; de Freitas Pacheco, J.A.; Idiart, T.P. 2000
	A\&AS, 145, 467
\bibitem[1991]{daco} Da Costa, G.S. 1991, 
        The Magellanic Clouds, IAU Symposium N$^{o}$ 148,
	R. Haynes and D. Milne, Edts, p.183
\bibitem[1989]{denn} Dennefeld, M. 1989,
        Recent developments of Magellanic Clouds research, 
         K.S. de Boer, F. Spite, G. Stasinska Edts, p.107
\bibitem[1983]{denn2} Dennefeld, M. \& Stasinska, G. 1983,
	A\&A 118, 234
\bibitem[1990]{dop1} Dopita, M.A. \& Meatheringham, S.J. 1990,
	ApJ357, 140
\bibitem[1991]{dop2} Dopita, M.A. \& Meatheringham, S.J. 1991,
	ApJ 367, 115
\bibitem[1991]{dopSMP64} Dopita, M.A. \& Meatheringham, S.J, 1991,
	ApJ 374, L21
\bibitem[1984]{duf1} Dufour, R.J., 1984,
	Structure and evolution of the MC's, IAU Symposium  108,
	S. van den Bergh \& K.S. de Boer, Eds, p.353
\bibitem[1999]{Garn} Garnett, D.R. 1999,
	New views of the Magellanic Clouds, IAU Symposium N$^{o}$ 190,
	Y.H. Chu, N.B. Suntzeff, J.E. Hesser, D.A. Bohlender, Edts, p.266
\bibitem[1989]{grev} Grevesse, N. \& Anders, E. 1989, AIP Conf. Proc. 183,
	Cosmic Abundances of Matter, ed. C.J. Waddington, p. 1
\bibitem[1993]{groe} Groenewegen M.A.T. \& de Jong, T. 1993,
	A\&A 267, 410
\bibitem[1956]{heni} Henize, K.G. 1956,
	ApJS 2, 315 (LHA-Nxx)
\bibitem[1989]{hen1} Henry, R.B.C. 1989,
	MNRAS 241, 453
\bibitem[2000]{herw1} Herwig, F. 2000, 
	A\&A, 360, 952
\bibitem[2004]{herw2} Herwig, F. 2004, 
	ApJS 155, 651
\bibitem[2000]{hill1} Hill, V., Francois, P., Spite, M. et al. 2000,
        A\&AS 364, L19
\bibitem[1977]{iben1} Iben I. 1977,
	ApJ 217, 788
\bibitem[1978]{iben2} Iben I. \& Truran, J.W. 1978,
	ApJ 220, 980
\bibitem[1983]{iben3} Iben I. \& Renzini, A. 1983,
	ARA\&A 21, 271
\bibitem[1999]{jacci} Jacoby, G.H. \& Ciardullo R. 1999,
	ApJ 515, 169
\bibitem[(1978)]{kal1} Kaler, J. B. Iben, I., Jr. \& Becker, S. A. 1978,
	 ApJ 224, 63
\bibitem[1994]{king} Kingsburgh, R. L. \& Barlow, M. J. 1994,
	MNRAS 271, 257
\bibitem[1996]{leis2} Leisy, P. \& Dennefeld, M. 1996,
        A\&AS 116, 96 (Paper~I)
\bibitem[2000]{leis3} Leisy, P. \& Dennefeld, M. 2000, Rev. Mex. AA 9, 227 
\bibitem[1996]{mari1} Marigo, P., Bressan, A. \& Chiosi, C. 1996,
	A\&A 313, 545 
\bibitem[2001]{mari2} Marigo, P. 2001, 
	A\&A 370, 194
\bibitem[2003]{mari3} Marigo, P., Bernard-Salas, J., Pottasch, S.R. et al. 2003, 
	A\&A 409, 619
\bibitem[1991a]{mea1} Meatheringham, S.J. \& Dopita, M.A. 1991,
	ApJS 75, 407 (Paper a)
\bibitem[1991b]{mea2} Meatheringham, S.J. \& Dopita, M.A. 1991,
	ApJS 76, 1085 (Paper b)
\bibitem[1988]{meam} Meatheringham, S.J., Dopita, M.A., Ford, H.C. \& Webster, B.L. 1988, 
        ApJ 327, 651
\bibitem[2002]{methan} Metcalfe, T.S., Salaris, M., Winget, D.E. 2002,
	ApJ 573, 803
\bibitem[1972]{milm} Miller J.S. \& Matthews W.G. 1972,
	ApJ 172, 593
\bibitem[1988]{monk} Monk, D.J., Barlow, M.J. \& Clegg, R.E.S. 1988,
	MNRAS 234, 583
\bibitem[1985]{mor1} Morgan, D.H. \& Good, A.R. 1985,
	MNRAS 213, 491 (MGPN SMC xx)
\bibitem[1992]{mor2} Morgan, D.H. \& Good, A.R. 1992,
	A\&AS 92, 571 (MGPN LMC xx)
\bibitem[1994]{mor3} Morgan, D.H. 1994, 
	A\&AS 103, 235 (M94b-xx)
\bibitem[1995]{mor4} Morgan, D.H. 1995, 
	A\&AS 112, 445 (M95-x)
\bibitem[1981]{Nand} Nandy, K., Morgan,D.H., Willis, A.J. et al. 1981,
	MNRAS 196, 955
\bibitem[1978]{peim1} Peimbert, M. 1978,
	Planetary Nebulae, IAU Symposium  76,
	Terzian Y., edt. (Dordrecht:Reidel), p215
\bibitem[1984]{peim2} Peimbert, M. 1984,
	Structure and Evolution of the MC's, IAU Symposium N$^{o}$ 108,
	S. van den Bergh and K.S. de Boer, Edts, p.363
\bibitem[1985]{peim3} Peimbert, M. 1985,
	Rev. Mex. Astr. Astrof. 10, 125
\bibitem[(1981)]{ren1} Renzini, A. \& Voli, M. 1981,
	A\&A 94, 175
\bibitem[1990]{rata} Ratag, M.A. \& Pottasch, S.R. 1990,
	A\&A 227, 207
\bibitem[1978]{san1} Sanduleak, N., MacConnell, D.J. \& Phillip, A.G.D. 1978,
	PASP 90, 621 (SMP LMC (or SMC) xxx)
\bibitem[1991]{spit1} Spite, F. \& Spite, M. 1991,
	The Magellanic Clouds, IAU Symposium N$^{o}$ 148,
	R. Haynes and D. Milne, Edts, p.243
\bibitem[2000]{stang1} Stanghellini, L., Shaw, R.A., Balick, B. \& Blades, J.C. 2000 
	ApJL, 534, 167
\bibitem[(1989)]{swe1} Sweigart, A. V., Greggio, L. \& Renzini, A. 1989, 
	ApJS 69, 911
\bibitem[(1990)]{swe2} Sweigart, A. V., Greggio, L. \& Renzini, A. 1990, 
	ApJ 364, 527
\bibitem[1990]{vdb} Van den Bergh, S. 1999, 
        New views of the Magellanic Clouds, IAU Symposium N$^{o}$ 190,
	Y.H. Chu, N.B. Suntzeff, J.E. Hesser, D.A. Bohlender, Edts, p. 569 
\bibitem[1992]{vasi1} Vassiliadis, E., Meatheringham, S.J. \& Dopita, M.A. 1992,
	A.J. 394, 489
\bibitem[1993]{vasi2} Vassiliadis, E. \& Wood, P.R. 1993
	ApJ 413, 641
\bibitem[2004]{wern1} Werner, K., Rauch, T., Reiff, E. et al. 2004,  
	A\&A 427, 685
\bibitem[2005]{wern2} Werner, K. \& Herwig, F. 2005, 
	Astro-ph 0512320 
\bibitem[1958]{whit} Whitford, A.E. 1958,
	A.J., 63, 201

\end{thebibliography}
\end{document}